\documentclass[aps,prl,twocolumn,superscriptaddress]{revtex4-1}

\usepackage{hyperref} 
\usepackage{graphicx}
\usepackage{amsmath}
\usepackage{amssymb}

\DeclareGraphicsExtensions{.png,.jpg,.eps}
\usepackage{xcolor}

\def\bk{{\bold{k}}}
\def\br{{\bold{r}}}

\def\bg{{\bold{g}}}

\def\m1{{^{-1}}}

\def\bias{U}
\def\mass{m}

\begin{document}

\author{A. Ramires}
\affiliation{Institute for Theoretical Studies, ETH Zurich, 8092 Zurich, Switzerland}

\author{J. L. Lado}
\affiliation{Institute for Theoretical Physics, ETH Zurich, 8093 Zurich, Switzerland}

\title{
Electrically tunable gauge fields 
in tiny-angle twisted bilayer graphene
}

\begin{abstract}
Twisted bilayer graphene has recently attracted a lot of attention for its rich electronic properties and tunability. 
Here we show that for very small twist angles, $\alpha \ll 1^\circ$, the application of a perpendicular
electric field is mathematically equivalent to a new kind of artificial gauge field. This identification opens the door for the generation and detection of pseudo-Landau levels in graphene platforms within robust setups which do not depend on strain engineering and therefore can be realistically harvested for technological applications. Furthermore, this new artificial gauge field leads to the development of highly localized modes associated with flat bands close to charge neutrality which form an emergent Kagome lattice in real space. Our findings indicate that  for tiny angles, biased twisted bilayer graphene is a promising platform which can realize frustrated lattices of highly localized states, opening a new direction for the investigation of strongly correlated phases of matter.

\end{abstract}

\date{\today}

\maketitle


Graphene is one of the most versatile materials to realize
exotic phenomena in condensed matter as a consequence of its emergent Dirac-like dispersion \cite{RevModPhys.81.109,novoselov2007room,zhang2005experimental}.
Yet, the presence of a Dirac point represents a big drawback if one is interested
in ordered states of matter: the density of states (DOS) at the Fermi energy is
zero, suppressing the development of electronic instabilities in
pristine graphene.  Such limitation is lifted in the presence of gauge fields
\cite{yang2007spontaneous}. In particular, under external magnetic fields, the
development of Landau levels yield a dramatic enhancement of the DOS at charge
neutrality, leading
to fractional quantum Hall states \cite{bolotin2009observation},
antiferromagnetism \cite{young2014tunable},  spin and valley ferromagnetism
\cite{young2012spin}, and spin superfluids
\cite{PhysRevLett.116.216801,stepanov2018long}.

In this context, artificial
gauge fields seem to be a promising direction for further investigation. 
On one hand, artificial gauge fields
would provide a new way to control the electronic
structure of graphene, such as emulating the
physics of extreme high magnetic fields \cite{Levy2010}.
On the other hand, the emergence of flat
pseudo Landau levels can realize frustrated Mott
insulators \cite{CaoMott2018},
potentially yielding 
quantum spin liquid states \cite{RevModPhys.88.041002}.
Interestingly,
the tunability of the electronic density and exchange interactions
of graphene provides a highly attractive solid state platform
where quantum spin liquid states could be electrically
tuned \cite{eich2018coupled} and even doped \cite{PhysRevX.6.041007},
in comparison with the limitations of natural compounds.
\cite{Banerjee2016,PhysRevB.91.144420,PhysRevX.6.041007}.
In this line, non-uniform strain is known to generate
artificial gauge fields \cite{Guinea2009,Voz2010},
yet a controllable and systematic realization may be
experimentally challenging \cite{Levy2010,amorim2016novel}. 
Therefore, the search for controllable ways to generate gauge fields in
graphene platforms is of high interest as they potentially provide
solid state realizations of exotic states of matter.

Introducing one extra layer of complexity, 
twisted bilayer graphene (TBG) is known to host a plethora of new interesting phenomena
\cite{ROZHKOV20161,PhysRevB.92.075402,PhysRevB.87.205429,wright2011robust,PhysRevX.3.021018,PhysRevB.89.205414,van2015relaxation,PhysRevB.91.155428,gargiulo2017structural,fleischmann2018moir,PhysRevMaterials.2.034004},
stemming from the appearance of a new scale: the Moir\'e
length $L_M$
\cite{PhysRevLett.107.216602,PhysRevLett.108.216802,PhysRevLett.119.107201,esquinazi2014superconductivity,huang2017evolution,bistritzer2011moire,PhysRevLett.117.116804}. 
 The physics of TBG strongly depends on the angle $\alpha$ between the two layers.
At angles $\alpha \gg 1 ^\circ$, the effect
of the twist is to renormalize the Fermi
velocity \cite{PhysRevB.86.155449,finocchiaro2017quantum}. 
For very small angles, $\alpha \ll 1^\circ$,
graphene superlattices are known to give rise to delocalized states forming a helical network when an electric bias is applied \cite{PhysRevB.88.121408,rickhaus2018transport,huang2018emergence}.
In the neighborhood of the magic angle $\alpha \approx 1^\circ$, the Fermi velocity
is heavily suppressed, giving rise
an almost flat
band \cite{PhysRevB.82.121407,PhysRevB.86.125413,kim2017tunable}. In this last regime, recent breakthrough
experiments have shown Mott insulating regime \cite{CaoMott2018} and
superconductivity \cite{cao2018unconventional}.
Unfortunately, these phenomena only occur at 
very specific magic angles, requiring a precisely tuned structure. 
It would be highly desirable to realize a similar situation
in a less fine tuned regime, preferably within setups which do not depend on lattice manipulations, but more simply on the application of electric biases.

In this letter, we analytically show that for TBG in the
tiny angle regime ($\alpha \ll 1^\circ$),
a homogeneous interlayer bias can be mapped into an artificial gauge field
whose magnitude is proportional to the applied electric bias. 
We corroborate  this result with exact numerical calculations which show the formation of a
discrete set of highly localized levels forming flat bands close to the charge neutrality point, in addition to the already established helical network states \cite{PhysRevB.88.121408}. The origin of these new localized states can be ascribed to the emergent gauge field and thus these are called
pseudo Landau levels (pLL).
Interestingly, these states arise
for a continuous
set of angles in presence of nonzero electrical bias, and do not require fine tuning to the magic angles. 
Furthermore, the lowest energy pLL forms an emergent Kagome lattice in real
space, making
TBG an attractive platform to realize quantum spin liquid states.

\begin{figure}[t!]
 \centering
                \includegraphics[width=\columnwidth]{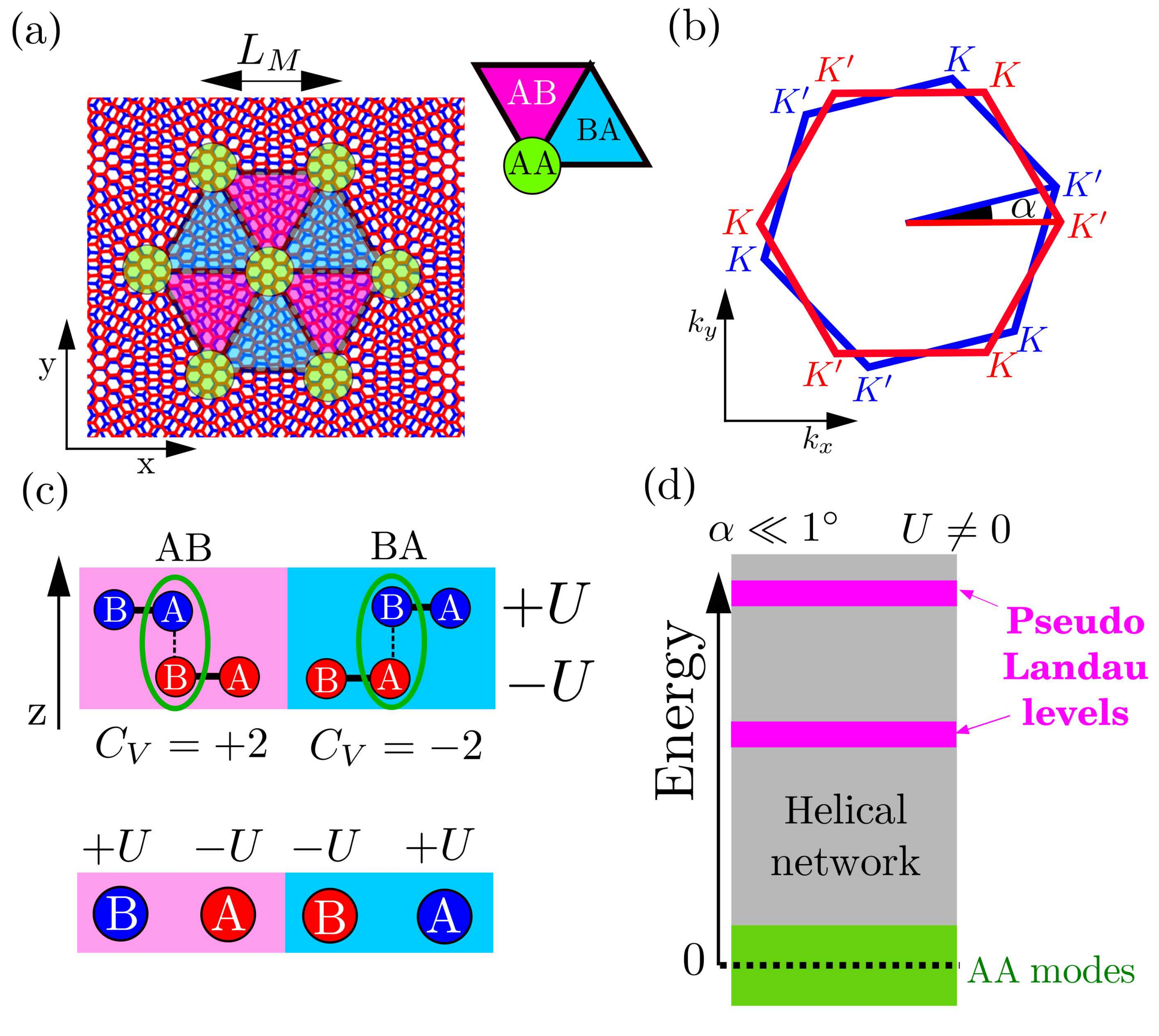}
                
\caption{
(a) Moir\'{e} pattern and the identification of regions with AB, BA and AA
stacking. (b) Brillouin zones of the lower and upper layers rotated by the
twisting angle $\alpha$. (c) Top: schematic representation  of a side view of
twisted bilayer graphene under external bias. We depict a set of representative sites in the AB (pink) and BA (blue) regions, associated with opposite valley Chern number
$C_V$. Bottom: effective low energy model
obtained by integrating out the dimers.
(d) Sketch of the spectra of tiny angle
($\alpha \ll 1^\circ$)
 twisted bilayer graphene under
external bias, showing states in the AA regions
at the Dirac point, helical networks in a wide energy
range, and electrically generated pseudo
Landau levels.
}
\label{fig1}
\end{figure}


TBG displays alternating patterns between AB and BA stacking in real space, together with zones with AA stacking (XY stacking corresponds to perfect alignment of a site from sublattice X in the upper layer with a site from sublattice Y from the lower layer), as shown schematically in Fig. \ref{fig1}a. 
In momentum space, the twist creates a relative rotation between the Brillouin zones (BZs) of the two layers, generating an effective mini-BZ (Fig. \ref{fig1}b).  
Regions with AB or BA stacking are associated with opposite valley Chern number $C_V$ (Fig. \ref{fig1}c), therefore one expects the emergence of low energy (topological) states at these interfaces. In the following we focus on deriving a low energy effective model for the regions of transition between AB and BA stacking. 
To this end, we start with the Hamiltonian for perfectly AB stacked bilayer graphene at the K-point. The dimerization between the atoms in sublattice A in the upper layer with the atoms in sublattice B in the lower layer (schematically shown in Fig. \ref{fig1}c ) gives rise to an effective two-component low energy model (valid for energies lower than the inter-layer coupling $t_\perp$) with quadratic band touching at
the K points, corresponding to massive chiral quasiparticles \cite{mccann2013electronic}.
The first component refers to sublattice B in the upper layer, whereas the
second component refers to sublattice A in the lower layer. In presence of an interlayer bias, the two components acquire different onsite energies, giving rise to the effective model
\begin{eqnarray}\label{H0}
H_0 (\bk) =  \gamma \boldsymbol{\sigma} \cdot \boldsymbol{\Gamma}_\bk + U \sigma_z,
\end{eqnarray}
where $ \boldsymbol{\sigma} = (\sigma_x,\sigma_y)$ and $\boldsymbol{\Gamma}_\bk
= (k_x^2-k_y^2, 2 k_xk_y)$. Here $\gamma$ gives a measure of the inverse
effective mass and $U$ is the strength of the interlayer bias. 
This model can be easily solved, leading to eigenstates $\langle \Psi_+ (\bk) |=  (\gamma k_+^2,  E - \bias)/N_0$ and $\langle \Psi_- (\bk) |  =  (-E + \bias , \gamma k_-^2)/N_0$, where $k_\pm = k_x \pm i  k_y$. The normalization constant is $N_0 = [(E - \bias)^2 + \gamma^2 k_p^4]^{1/2}$, where $k_p^2 = k_x^2 + k_y^2$. The respective eigenenergies are $E_{\pm} = \pm E =  \pm (\bias^2 + \gamma^2 k_p^4)^{1/2}$. Note
that as soon as $\bias\neq 0$ a gap opens in the spectrum, which is symmetric
around zero energy.

We now extend this model to the case of very small angle TBG. First, it is interesting to note that the AB/BA interface with uniform bias is topologically equivalent to a uniform
AB stacking with alternating bias. This can be understood as the change in labels across the AB/BA boundary depicted in Fig. \ref{fig1}c. This picture can be made
concrete by performing a smooth spatially dependent unitary transformation
$R(\br)$ on the effective Hamiltonian in Eq. \ref{H0}, defining the modulated
Hamiltonian $H_M(\br) = R(\br) H_0(\br) R^\dagger (\br)$, with $R(\br)=I$ and
$\sigma_x$ for AB and BA regions, respectively. 
In the new rotated basis, $H_M(\br)$ shows a spatially 
dependent interlayer bias $\bias(\br)$ and reads:
\begin{eqnarray}
H_M(\br) =  \gamma (\boldsymbol{\sigma}\cdot \boldsymbol{\Gamma}) +
\bias(\br)\sigma_z,
\end{eqnarray}
where $\boldsymbol{\Gamma} = (\partial_x^2-\partial_y^2, \partial_x\partial_y+
\partial_y\partial_x)$.

The solution for the modulated problem in the rotated frame
can be found building on the solution of the unmodulated
case \cite{PhysRevLett.105.156801, ando2005theory}. Given the implicit
dependence of the eingenstates on the interlayer bias, $|\Psi_\pm (\br)\rangle =
|\Psi_\pm (\br, U)\rangle $, 
one can write a generalized eigenvalue problem associating the spatial dependence of the
interlayer bias with an independent coordinate
$
H_G(\br ' ,\br) | \Psi (\br', \br) \rangle = \epsilon| \Psi (\br', \br)\rangle,
$
where the generalized Hamiltonian reads
\begin{eqnarray}
H_G(\br ' ,\br)  =  \gamma (\boldsymbol{\sigma}\cdot \boldsymbol{\Gamma}') +  \gamma (\boldsymbol{\sigma}\cdot \boldsymbol{\Gamma})+ \bias(\br)\sigma_z,
\end{eqnarray}
with $\br'$ corresponding to the original coordinates and $\br$ corresponding
to the implicit dependence through $\bias(\br)$. The first term takes care of
the derivatives with respect to the explicit spatial coordinates, while the
second term takes care of the derivatives of the coordinates implicit in
$\bias(\br)$. In the limit $\br ' \rightarrow \br$, we recover the problem we
actually want to solve \cite{PhysRevLett.105.156801, ando2005theory}.

For small bias strength, meaning $U < t_\perp$, such that the low energy effective model is valid, the eigenstates should be similar to the ones
obtained for the unmodulated solution, therefore we consider an ansatz of the
form 
\begin{eqnarray}
|\Psi (\br', \br) \rangle= f_+(\br) |\Psi_+ (\br', \br) \rangle+ f_-(\br) |\Psi_- (\br', \br)\rangle,
\end{eqnarray}
where $|\Psi_\pm (\br', \br) \rangle = |\Psi_\pm (\br', \bias(\br))\rangle$ and
$f_\pm(\br)$ are smooth functions to be determined.
The generalized eigenvalue problem can then be explicitly written as:
\begin{eqnarray}\label{EigenV}
[\gamma\boldsymbol{\sigma}\cdot \boldsymbol{\Gamma} + (E_+(\br) - \epsilon)] f_+(\br) |\Psi_+ (\br', \br)\rangle
\\ \nonumber 
+ (f_+, \Psi_+, E_+ \rightarrow f_-, \Psi_-,E_-) = 0,
\end{eqnarray}
after simplification using the solution of the unmodulated problem.  In order to determine the functions $f_\pm(\br)$, we can evaluate the matrix
elements of the equation above with $|\Psi_\pm (\br', \br)\rangle$. We define
the matrix elements as
\begin{eqnarray}
\langle \Psi_1 | A | \Psi_2 \rangle_\br &=& \int d\br '  g(\br'-\br) \langle
\Psi_1^\dagger (\br',\br)|  A | \Psi_2(\br',\br)\rangle \\ \nonumber
&\approx&  \int d\bk  \langle \Psi_1^\dagger (\bk,\br)|  A   | \Psi_2(\bk,\br)\rangle,
\end{eqnarray}
where $g(\br-\br')$ is a smooth function peaked at zero which integrates to unit \cite{ando2005theory, PhysRevLett.105.156801}. Here  $| \Psi(\bk,\br)\rangle$ is the Fourier transform of $| \Psi(\br ',\br)\rangle$ with respect to its first spatial variable.
The only non-zero matrix elements are identified as:
\begin{eqnarray}\label{MatEl}
\langle \Psi_\pm |\boldsymbol{\sigma}| \Psi_\mp \rangle_\br  &=&  (v_x(\br), i
v_y(\br)),  \\
\langle \Psi_\pm |\boldsymbol{\sigma} \partial_U| \Psi_\mp \rangle_\br  &=& (
a_x(\br) ,  i a_y(\br) ),
\\
\langle \Psi_\pm |\boldsymbol{\sigma} \partial_U^2 | \Psi_\mp \rangle_\br  &=&
(b_x(\br),  i b_y(\br)).
\end{eqnarray}
The explicit forms of $v_{x,y}(\br)$, $a_{x,y}(\br)$ and $b_{x,y}(\br)$ are given in the Supplemental Material (SM). 
The eigenvalue equation now becomes:
\begin{eqnarray}\label{D2}
\begin{pmatrix}
E(\br) & \Pi(\br)\\
\Pi^*(\br) & -E(\br)
\end{pmatrix}\begin{pmatrix}
f_+(\br)\\
f_-(\br)
\end{pmatrix} = \epsilon \begin{pmatrix}
f_+(\br)\\
f_-(\br)
\end{pmatrix},
\end{eqnarray}
where
\begin{eqnarray}
\Pi(\br) &=& \gamma_x(\br) \Gamma_x + i \gamma_y(\br) \Gamma_y \\ \nonumber
&+& A_x(\br)
\partial_x +  A_y(\br) \partial_y  + C(\br),
\end{eqnarray}
with $\gamma_{x,y}(\br)  = \gamma v_{x,y}(\br)$ and
\begin{eqnarray}
A_{x}(\br)  &=& 2 \gamma a_{x}(\br) \partial_x \bias(\br) + i \gamma a_y (\br) \partial_y \bias(\br)\\ \nonumber
A_{y}(\br)  &=& -2 \gamma a_{x}(\br) \partial_y \bias(\br) + i \gamma a_y (\br) \partial_x \bias(\br).
\end{eqnarray}
The explicit form of $C(\br)$ is given in the SM.


The presence of gauge fields becomes evident
if we consider the following Peierls-like substitution $\partial_{x,y}
\rightarrow \partial_{x,y} + g_{x,y}(\br)$ for a generic two-component gauge
field $g_{x,y}(\br)$, in which case we can identify
\begin{eqnarray}
g_x(\br) &=&\frac{A_x(\br) \gamma_x(\br)- i A_y(\br)\gamma_y(\br)}{2(\gamma_x^2(\br) + \gamma_y^2(\br))},\\  
g_y(\br) &=&\frac{A_y(\br) \gamma_x(\br)- i A_x(\br)\gamma_y(\br)}{2(\gamma_x^2(\br) + \gamma_y^2(\br))},
\end{eqnarray}
as gauge fields induced by the modulation of the electric bias in
TBG graphene.
Note that these are given in terms of $A_{x,y}(\br)$ and $\gamma_{x,y}(\br)$,
which in turn are determined in terms of matrix elements defined above and the derivatives of the bias.
Note that the gauge fields are maximal where the derivatives of the bias have an extreme, meaning, the gauge fields are the largest in the AB/BA interfaces. This suggests that low energy modes are going to be localized at these interfaces. Finally, it is interesting to note that a similar treatment is also applicable for single or trilayer graphene on top of boron nitride (more details on the single layer scenario are provided in section III of the SM, which includes Refs. \cite{Yankowitz2012, Chen2018}).

To asses the validity of the mapping
above, we now perform an exact numerical evaluation of the states in twisted
bilayer graphene in presence of a homogeneous interlayer bias. 
We use a real space
tight binding Hamiltonian for twisted bilayer graphene \cite{PhysRevB.92.075402}
of the form
\begin{equation}
H = 
t
\sum_{\langle ij \rangle }
c^\dagger_i c_j + 
\sum_{ij}
\hat t_\perp(\br_i,\br_j) c^\dagger_i c_j
	+ U \sum_i \tau_z^{ii}
c^\dagger_i c_i ,
\label{htb}
\end{equation}
with $t$
the first neighbor hopping,
$\langle ij\rangle$ the
sum over first neighbors,
$\hat t_{\perp}(\br_i,\br_j)$
the distance dependent inter-layer coupling
taking a maximum value $t_\perp$ for perfect
stacking,
$U$ the interlayer bias,
and $\tau_z^{ii}=\pm 1$ labels the upper/lower layer.
As a reference, the
values of the parameters in graphene are
$t\approx 3$ eV and $t_\perp \approx 300$ meV \cite{mccann2013electronic}. More details on the computational aspects can be found in section IV of the SM, which also includes Refs. \cite{PhysRevLett.109.196802, RevModPhys.78.275}. 
With this tight binding Hamiltonian,
we numerically calculate the electronic spectra
close to the charge neutrality point
for different twisting angles and interlayer bias.

\begin{figure}[t!]
 \centering
                \includegraphics[width=\columnwidth]{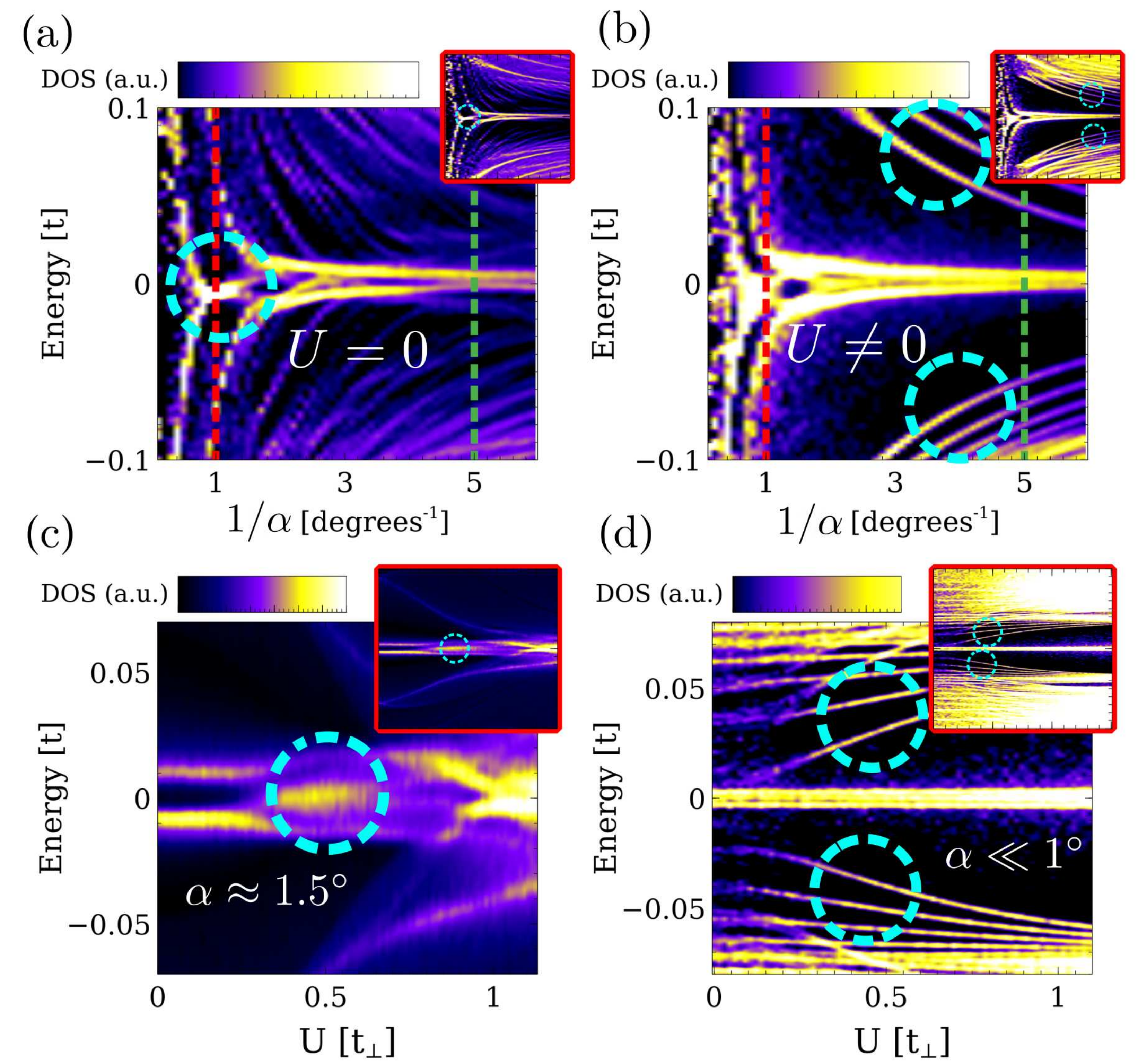}
                
\caption{ 
Evolution of the density of states as a function of the
twisting angle $\alpha$ and bias $U$. (a) Without bias ($U=0$), a peak at the magic angle
$\alpha \approx 1^\circ$ appears, highlighted by the cyan circle. (b) In
presence of interlayer bias ($U=0.5t_{\perp}$), a set of peaks develop  at very
small angles $\alpha\approx 0.3^\circ$, highlighted by cyan circles. Dashed
lines in (a) and (b) mark $\alpha=1^\circ$ (red) and $\alpha=0.2^\circ$
(green).
Evolution of the density of states as a function of
the electric bias for $\alpha \approx 1.5 ^\circ$ (c) and  $\alpha \approx 0.2 ^\circ$ (d).
For large angles, the electric bias enhances the DOS at zero energy (cyan circle in (c)); whereas for small angles a new set of pLL resonances emerge (cyan circles in (d)).
Bias in (c,d) is measured in units of the interlayer hopping $t_{\perp}$.
Insets show the DOS in a wider energy range.
}
\label{fig2}
\end{figure}

First, it is instructive to explore the electronic properties in
the absence of interlayer bias as a function
of the twisting angle between the two layers (Fig. \ref{fig2}a).
It is observed that at the magic angle $\alpha \approx 1 ^\circ$, a divergent
DOS arises at half filling, but it disappears as one slightly departs from that
angle. As the angle becomes much smaller than $1 ^\circ$, more
states start to flow towards zero energy,
creating a complex background of states.
The electronic structure as a function of $\alpha$
becomes much cleaner when the interlayer bias is switched on (Fig. \ref{fig2}b).
In particular, in the vicinity of zero energy
the spectrum develops
a smooth background with a very low density of states. This corresponds to the helical network
of states between AB and BA regions \cite{PhysRevB.88.121408,rickhaus2018transport}. 
The most interesting feature is the appearance of a set
of sharp resonances 
as shown in Fig. \ref{fig2}b.
Given the analytic results presented above,
these resonances can be associated with spatially localized modes, or pLL, stemming from the bias induced gauge field (further discussion can be found in section V of the SM, which also includes Refs. \cite{PhysRevB.90.155415, PhysRevLett.120.086603}).
Interestingly, these $U\ne0$ 
resonances exist for every angle $\alpha  \ll 1^\circ$
in contrast with the localized states at the magic angle resonance in Fig. \ref{fig2}a.

Now we explore the evolution of these localized modes
as a function of the interlayer bias. It is
important to point out that the behavior between
small $\alpha \approx 1.5 ^ \circ$ and very small
angles $\alpha \ll 1 ^ \circ$ is radically different.
For $\alpha \approx 1.5^ \circ $, the effect of the interlayer bias is
to enhance the DOS at charge neutrality, without creating
new resonances at nearby energies
(Fig. \ref{fig2}c)\cite{PhysRevLett.119.107201}.
In striking contrast, for $\alpha \ll 1^ \circ$,
the interlayer bias $U$
creates the completely new set
of pseudo Landau levels,
whose energy increases with the interlayer bias
in analogy with the magnetic field dependence of actual 
Landau levels (Fig. \ref{fig2}d).

\begin{figure}[t!]
 \centering
                \includegraphics[width=\columnwidth]{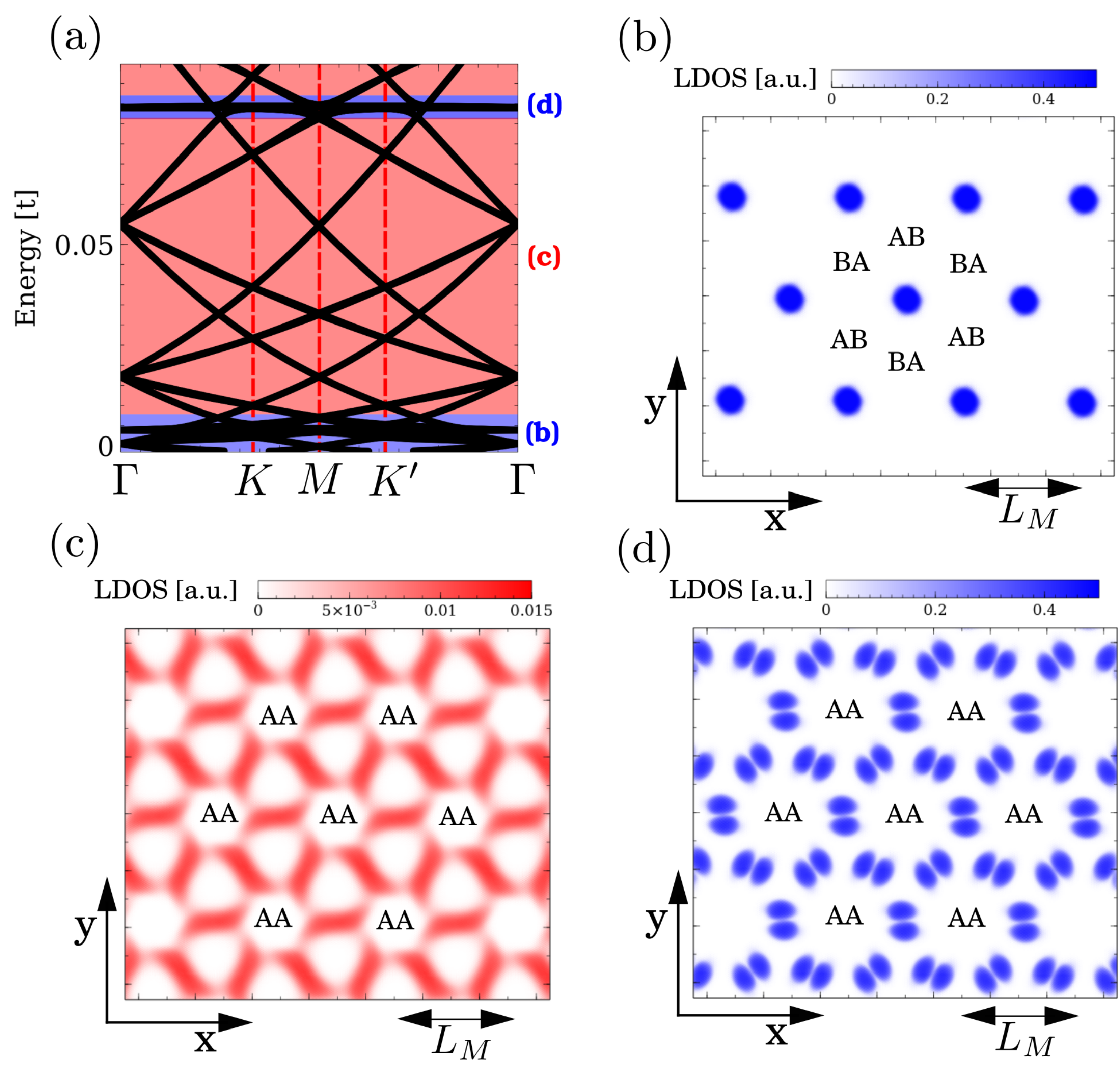}

\caption{ 
(a) Band structure in the pseudo Landau level regime
($\alpha \ll 1^\circ$, $U\ne0$), showing three groups of states: AA
states (b), helical networks (c) and pseudo Landau levels (d). The
calculation of the spatial distribution of the states show that (b) and
(d) are localized, whereas the helical network is delocalized (c).
Interestingly, states in (d) form a Kagome lattice, in comparison with the
triangular lattice formed by (b).
The lattice constant of the emergent lattices in (b,d)
is the Moir\'e length $L_M$.
}
\label{fig3}
\end{figure}

Further insight can be gained by observing the band-structure
for $\alpha \ll 1^ \circ$ and $U\ne0$ (Fig. \ref{fig3}a).
In particular, we observe that the bright resonance at charge neutrality
corresponds to a large set of nearly flat bands, 
whose wavefunction is localized
in the AA regions of TBG (Fig. \ref{fig3}b) \cite{PhysRevB.86.125413}.
Away from charge neutrality, a 
continuum of
delocalized highly dispersive states
shows up, whose spatial profile is between the
AB and BA regions (Fig. \ref{fig3}c) and correspond to the so
called helical network states.
Finally, at slightly higher energy a set of flat bands appear, which corresponds to one of the
pLL resonances highlighted Figs. \ref{fig2}b,d.
The spatial profile of the pLL flat bands
is also between the AB and BA regions, but contrary
to the helical modes, it remains strongly localized.
A last interesting note comes from the fact that these localized modes
form an emergent Kagome lattice (Fig. \ref{fig3}d), and the band counting yields 2 states
per site (6 spinless modes per Moir\'e unit cell). 
This suggests that twisted bilayer graphene is a potential tunable solid state platform for frustrated magnetism and quantum spin liquid physics, as Kagome lattices are known to be ideal playgrounds to realize such highly entangled ground states \cite{anderson1973resonating,PhysRevX.6.041007,RevModPhys.88.041002}.

A final remark concerns the regime of validity of the analytic calculation,
so that its comparison with the numerical results remains meaningful.
First, the two-dimensional effective model for AB graphene is valid for
energies smaller than the inter-layer bias, which in turn should be smaller than the inter-layer coupling $E<U<t_\perp$.
Moreover, given that we are working
with a continuum model starting from Bernal stacking, 
the approximation is valid for very small $\alpha$,
i.e. big AB regions. As a consequence, we estimate that 
an experimentally realistic regime for the observation of the pLL peaks 
would require $\alpha\approx 0.1-0.3^\circ$ and $U\approx100-200$ meV, so that
the localized modes would arise at energies around $20-50$ meV
with respect to the charge neutrality point. 
Interestingly, the requirements of our proposal
have been almost already fufilled in
recent experiments \cite{rickhaus2018transport},
which will potentially allow
to directly characterize pseudo-Landau levels by transport
measurements in a single electronic device.

To summarize, we have established both analytically and
numerically that an electric bias
in TBG creates an artificial gauge field.
We have identified the bias induced pseudo Landau levels
by exactly solving a tight binding Hamiltonian for TBG,
and we have shown that these levels are located close to charge neutrality,
implying that they can be easily accessed in a realistic
experimental situation.
Moreover, we have shown that the emergent resonances correspond
to flat bands whose spatial distribution creates an emergent Kagome lattice in graphene.
Importantly, these localized modes exist for a continuum  set of angles below
the magic angle, avoiding the fine tuning requirement for a flat band at the
magic angle. Our findings open a new direction for the investigation of the
physics of strong
correlations in graphene, which will certainly bring us new insights concerning exotic
phases of matter such as unconventional superconductivity and spin-liquid
behavior.

\textbf{Acknowledgments}
We would like to thank W. Chen, O. Zilberberg , J. Fernandez-Rossier and P. Rickhaus
for helpful discussions.
A.R.
is grateful for financial support by Dr. Max R\"ossler, the
Walter Haefner Foundation and the ETH Zurich Foundation.
J.L.L is grateful for financial support from ETH
Fellowship program
and from the
JSPS Core-to-Core program ``Oxide Superspin" international network.


\section*{Supplemental Material}

\appendix

In this Supplemental Material we give more details on the analytic and
numerical calculations outlined in the main manuscript.

\section{Explicit form of the matrix elements}

The first non-zero matrix element is:
\begin{eqnarray}
\langle \Psi_\pm |\boldsymbol{\sigma}| \Psi_\mp \rangle_\br  &=&  (v_x(\br), i v_y(\br)),
\end{eqnarray}
with
\begin{eqnarray}
v_{x,y}(\br) = - \! \int \!\! \frac{d\bk}{(2\pi)^2}\frac{[E(\br)-\bias(\br)]^2 \pm\gamma^2 k_+^4 }{N_0^2(\br)} ,
\end{eqnarray}
where the $+$ stands for $v_{x}(\br)$ and the $-$ for $v_{y}(\br)$.

The second non-zero matrix element involves a first derivative:
\begin{eqnarray}
\langle \Psi_\pm |\boldsymbol{\sigma} \partial_\bias| \Psi_\mp \rangle_\br  &=& ( a_x(\br) ,  i a_y(\br) ),
\end{eqnarray}
with
\begin{eqnarray}
a_{x,y}(\br) &=& \int \frac{d\bk}{(2\pi)^2} \Bigg\{ \pm F(\br) \gamma^2(k_x^4 - 6 k_x^2k_y^2 + k_y^4)\\ \nonumber
&+&  [E(\br)-\bias(\br)]^2 \left[\frac{1}{N_0^2(\br) E(\br)}  - F(\br)\right]  \Bigg\},
\end{eqnarray}
where again the $+$ sign stands for $a_{x}(\br)$ and the $-$ sign for $a_{y}(\br)$ and
\begin{eqnarray}
F(\br) = \frac{1}{N_0(\br)} \partial_{\bias} \left(\frac{1}{N_0(\br)}\right).
\end{eqnarray}

The third non-zero matrix element is particular to TBL graphene since it
carries a second derivative as follows:
\begin{eqnarray}
\langle \Psi_\pm |\boldsymbol{\sigma} \partial_\bias^2 | \Psi_\mp \rangle_\br  &=&  (b_x(\br),  i b_y(\br)),
\end{eqnarray}
with
\begin{eqnarray}
b_{x,y}(\br) &=& \int \frac{d\bk}{(2\pi)^2} \Big[ G(\br)v_{x,y}(\br) + 2 F(\br) \frac{[E(\br)-\bias(\br)]^2}{E(\br)} 
\nonumber \\ 
&-&\frac{[E^2(\br)-\bias^2(\br)](E(\br)-\bias(\br))}{N_0^2(\br)E^3(\br)}   \Big],
\end{eqnarray}
where
\begin{eqnarray}
G(\br) &=& \partial_{\bias} \left[N_0 (\br)\partial_{\bias}
\left(\frac{1}{N_0(\br)}\right)\right] \\ \nonumber
&+& N_0^2(\br) \left[\partial_{\bias} \left(\frac{1}{N_0(\br)}\right)\right]^2.
\end{eqnarray}


\section{Identification of the gauge field for twisted bilayer}

The presence of gauge fields becomes evident if we consider the following Peierls-like substitution for a generic gauge field:
\begin{eqnarray}
\partial_{x,y} \rightarrow \partial_{x,y} + g_{x,y}(\br),
\end{eqnarray}
in which case the derivative terms in the Hamiltonian transform as:
\begin{eqnarray}
\gamma_x(\partial_{x}^2 -\partial_y^2) &+& i \gamma_y (\partial_x\partial_y + \partial_y\partial_x) \rightarrow \\ \nonumber
&&\gamma_x(  \partial_x^2 -\partial_y^2) +i  \gamma_y (  \partial_x\partial_y+ \partial_y\partial_x) \\ \nonumber
&&+ 2 [ \gamma_x g_x(\br) +i  \gamma_y g_y(\br)]\partial_x \\ \nonumber
&&+ 2 [ \gamma_x g_y(\br) +i  \gamma_y g_x(\br)]\partial_y + cte,
\end{eqnarray}
so we can identify:
\begin{eqnarray}
A_x(\br) &=& 2 [ \gamma_x g_x(\br) +i  \gamma_y g_y(\br)], \\ \nonumber
A_y(\br) &=& 2 [ \gamma_x g_y(\br) +i  \gamma_y g_x(\br)], \\ \nonumber
C(\br)&=& (g_x(\br) + g_y(\br))^2 + (\partial_x g_x(\br)) + (\partial_y g_y(\br))\\ \nonumber 
&+& (\partial_y g_x(\br)) + (\partial_x g_y(\br)),
\end{eqnarray}
and solve for the unknown gauge fields:
\begin{eqnarray}
g_x(\br) &=&\frac{A_x(\br) \gamma_x(\br)- i A_y(\br)\gamma_y(\br)}{2(\gamma_x^2(\br) + \gamma_y^2(\br))},\\  
g_y(\br) &=&\frac{A_y(\br) \gamma_x(\br)- i A_x(\br)\gamma_y(\br)}{2(\gamma_x^2(\br) + \gamma_y^2(\br))}.
\end{eqnarray}


\section{Artificial gauge fields in single layer graphene with modulated sublattice imbalance}

The results in the main text were obtained for twisted bilayer graphene, but
similar results are valid for single or triple layer graphene with effective models including a modulated
sublattice imbalance. This scenario is experimentally relevant considering the
observation of Moir\'{e} patterns in single and triple layer graphene on top of
boron nitride \cite{Yankowitz2012,Chen2018}. In this appendix we develop the
derivation for single layer graphene, in which case the gauge fields take a
simpler form than in the main text. The derivation for triple-layer is also
possible and would follow similar lines. Here we use the same notation
introduced in the main text, but this section should be seen as self-contained.

The low energy effective Hamiltonian for graphene with a constant sublattice
imbalance can be written, near the $K$ point, as:
\begin{eqnarray}\label{H0SM}
H_0 (\bk) = \begin{pmatrix}
\mass & \gamma k_-\\
\gamma k_+ & -\mass
\end{pmatrix},
\end{eqnarray}
where $k_\pm = k_x \pm i k_y$, or more concisely:
\begin{eqnarray}
H_0 (\bk) =  \gamma \boldsymbol{\sigma} \cdot \bk + \mass \sigma_z,
\end{eqnarray}
where $ \boldsymbol{\sigma} = (\sigma_x,\sigma_y)$, $\bk = (k_x,k_y)$ and $\mass$
is the strength of the sublattice imbalance. Now the Hamiltonian is written in
the sublattice basis for a single graphene sheet. This model can be easily
solved, leading to eigenenergies $E_{\pm} = \pm E =  \pm \sqrt{\mass^2 +
\gamma^2 k_p^2}$, where $k_p^2 = k_x^2 + k_y^2$, and the respective
eigenstates:
\begin{eqnarray}\label{ES1}
|\Psi_+ (\bk) \rangle= \frac{1}{N_0}\begin{pmatrix}
\gamma k_-\\
E - \mass
\end{pmatrix},
\end{eqnarray}
\begin{eqnarray}\label{ES2}
|\Psi_- (\bk) \rangle = \frac{1}{N_0}\begin{pmatrix}
-E + \mass\\
\gamma k_+
\end{pmatrix},
\end{eqnarray}
with normalization constant $N_0 = \sqrt{(E - \mass)^2 + \gamma^2 k_p^2}$.

From the solution of the unmodulated problem, we can proceed as in the main
text in order to find the solution of the modulated case \cite{sun2010,
ando2005}. Writing the Hamiltonian in space coordinates and taking $\mass
\rightarrow \mass(\br)$:
\begin{eqnarray}
H_M(\br) =  -i  \gamma (\boldsymbol{\sigma}\cdot \nabla) +  \mass(\br)\sigma_z,
\end{eqnarray}
leads to the generalized eigenvalue problem:
\begin{eqnarray}
H_G(\br ' ,\br) \Psi (\br', \br) = \epsilon \Psi (\br', \br),
\end{eqnarray}
with
\begin{eqnarray}
H_G(\br ' ,\br)  =  -i  \gamma (\boldsymbol{\sigma}\cdot \nabla' + \boldsymbol{\sigma}\cdot \nabla) + \mass(\br)\sigma_z,
\end{eqnarray}
and
\begin{eqnarray}
\Psi_\pm (\br',\br) &=& \int d\bk \Psi_\pm (\bk, \mass(\br)) e^{i \bk \cdot
\br' } \\ \nonumber &=&  \int d\bk \Psi_\pm (\bk, \br) e^{i \bk \cdot \br' }.
\end{eqnarray}
As discussed in the main text, $\br'$ correspond to the original coordinates and $\br$ corresponds to the implicit dependence through $\mass(\br)$.

We look for solutions of the form:
\begin{eqnarray}
\Psi (\br', \br) = f_+(\br) \Psi_+ (\br', \br) + f_-(\br) \Psi_- (\br', \br),
\end{eqnarray}
where $\Psi_\pm (\br', \br) = \Psi_\pm (\br', \mass(\br))$ and $f_\pm(\br)$ are
smooth functions to be determined.   We are going to use the explicit form of
the solution $ \Psi_\pm (\br', \mass(\br))$ to find matrix elements of the
eigenvalue equation above in order to determine these functions.

Writing the generalized eigenvalue problem explicitly:
\begin{eqnarray}
&-&i  \gamma f_+(\br)  (\boldsymbol{\sigma}\cdot \nabla')  \Psi_+ (\br', \br) \nonumber
-i  \gamma f_+(\br)  (\boldsymbol{\sigma}\cdot \nabla)  \Psi_+ (\br', \br) \\ \nonumber
&-&i  \gamma   \boldsymbol{\sigma}\cdot (\nabla  f_+(\br)) \Psi_+ (\br', \br) 
+ \mass(\br)\sigma_z f_+(\br) \Psi_+ (\br', \br)  \\ 
&-& \epsilon f_+(\br) \Psi_+ (\br', \br) 
+ (f_+, \Psi_+ \rightarrow f_-, \Psi_-) = 0,
\end{eqnarray}
and simplifying, using the solution of the unmodulated problem:
\begin{eqnarray}
&-&i  \gamma f_+(\br)  (\boldsymbol{\sigma}\cdot \nabla)  \Psi_+ (\br', \br) \nonumber
-i  \gamma   \boldsymbol{\sigma}\cdot (\nabla  f_+(\br)) \Psi_+ (\br', \br)   \\ \nonumber
&+&(E_+(\br) - \epsilon) f_+(\br) \Psi_+ (\br', \br) \\ 
&+& (f_+, \Psi_+, E_+ \rightarrow f_-, \Psi_-,E_-) = 0.
\end{eqnarray}

We can now take matrix elements, as defined in the main text, to find the following non-zero terms:
\begin{eqnarray}
\langle \Psi_+   | \sigma_x | \Psi_-   \rangle_\br  
&=&  v (\br),
\end{eqnarray}
\begin{eqnarray}
\langle \Psi_+   | \sigma_y | \Psi_-   \rangle_\br 
&=&  i v(\br),
\end{eqnarray}
where
\begin{eqnarray}
v(\br) = - \int d \bk \frac{[E(\br)-\mass(\br)]^2}{N_0^2(\br)};
\end{eqnarray}
and
\begin{eqnarray}
\langle \Psi_-   | \sigma_x \partial_x | \Psi_+   \rangle_\br 
&=&  \frac{\partial  \mass(\br)}{\partial x} a(\br),
\end{eqnarray}
\begin{eqnarray}
\langle \Psi_-   | \sigma_y \partial_y | \Psi_+   \rangle_\br 
&=&  -i \frac{\partial  \mass(\br)}{\partial y} a(\br),
\end{eqnarray}
where
\begin{eqnarray}
a(\br) &=&   \int d \bk  \Bigg( - \frac{[E(\br)-\mass(\br)]^2}{N_0^2(\br) E(\br)}   \\ \nonumber 
&+&  \frac{[E(\br)-\mass(\br)]^2}{N_0^4(\br) E(\br)}\left\{ \gamma ^2 k_p^2
+[E(\br)-\mass(\br)]^2 \right\} \Bigg).
\end{eqnarray}

The eigenvalue problem for the functions $f_{\pm}(\br)$ can then be concisely written as:
\begin{eqnarray}
\begin{pmatrix}
E(\br) &  - i \gamma \Pi(\br)\\
- i \gamma \Pi^*(\br) & -E(\br)\\
\end{pmatrix}
\begin{pmatrix}
f_+(\br)\\
f_-(\br)
\end{pmatrix}
= \epsilon
\begin{pmatrix}
f_+(\br)\\
f_-(\br)
\end{pmatrix},
\end{eqnarray}
where
\begin{eqnarray}
\Pi(\br) &=& \left[ v(\br) \partial_x + a(\br)\frac{\partial
\mass(\br)}{\partial x}\right] \\ \nonumber&+& i \left[ v(\br) \partial_y -
a(\br)\frac{\partial \mass(\br)}{\partial y} \right],
\end{eqnarray}
from which we can directly identify the artificial gauge fields as:
\begin{eqnarray}
\bg (\br) = \frac{a(\br)}{v(\br)} \left( \frac{\partial
\mass(\br)}{\partial x}, - \frac{\partial
\mass(\br)}{\partial y}\right).
\end{eqnarray}
Note that the presence of the modulated sublattice imbalance
also leads to a renormalization of the effective velocity.

\section{Computational Details}
\subsection{Geometry of twisted bilayer graphene}

Unit cells of twisted bilayer graphene can be generated by taking into
account that the lattice vectors $\vec A_1, \vec A_2$ of a
commensurate Moir\'e supercell
follows \cite{PhysRevB.92.075402}:
\begin{eqnarray}
\vec A_1 &=& m_0 \vec a_1 + (m_0 + r) \vec a_2,\\
\vec A_2 &=& -(m_0+r) \vec a_1 + (2m_0 + r) \vec a_2,
\end{eqnarray}
with $m_0$ and $r$ integers
and $\vec a_1, \vec a_2$ the
graphene lattice vectors. The angle between the two layers in this supercell
is of the form \cite{PhysRevB.92.075402}:
\begin{equation}
\cos \alpha = 
\frac{3 m_0^2 + 3m_0 r + r^2/2}{3 m_0^2 + 3 m_0 r + r^2}.
\end{equation}

In our calculations, we take $r=1$, so
that the angle between the unit cells is controlled by $m_0$.
In that situation, the total number of atoms in the unit cell
is
\begin{equation}
N_C = 4(3m_0^2 + 3m_0 + 1).
\end{equation}

Given a certain $m_0$, the Moir\'e period as a function
of the twist angle $\alpha$ takes the form
\begin{equation}
L_M \propto \frac{1}{2\sin\frac{\alpha}{2}},
\end{equation}
which in the small angle limit becomes
\begin{equation}
L_M \propto \frac{1}{\alpha},
\end{equation}
and thus the size of the unit cell becomes bigger as the angle
$\alpha$ approaches zero.

\subsection{Tight binding Hamiltonian for twisted bilayer graphene}
Our numerical calculations are performed by numerically
solving a tight binding model for twisted bilayer
graphene of the form \cite{PhysRevB.92.075402,PhysRevLett.119.107201}

\begin{equation}
H = 
\sum_{\langle ij \rangle }
t c^\dagger_i c_j + 
\sum_{ij}
t_{\perp}(\vec r_i,\vec r_j) c^\dagger_i c_j
	+ U \sum_i \tau_z^{ii}
c^\dagger_i c_i ,
\label{htb}
\end{equation}
where $t=-2.7$ eV is the first neighbor hopping
and $\langle ij \rangle$ denotes sum over first neighbors within a layer.
The second term involving
$t_{\perp}(\vec r_i,\vec r_j)$ denotes the interlayer
hopping, that depends on the coordinates of the two atoms
$\vec r_i = (x_i,y_i,z_i)$
and 
$\vec r_j = (x_j,y_j,z_j)$ as

\begin{equation}
\hat t_{\perp}(\bold r_i,\bold r_j) = 
t_{\perp}
\frac{(z_i - z_j)^2 }{|\bold r_i - \bold r_j|^2}
e^{-\beta (|\bold r_i - \bold r_j|-d)},
\end{equation}
where $d$ is the interlayer distance
that for simplicity we take as $d=3a$, with
$a$ the carbon-carbon distance.
The parameter $t_{\perp}$
is the interlayer hopping when two carbon atoms sit one on
top of the other, which is the maximum value
of the interlayer coupling.
The parameter $\beta$ controls the decay of the
interlayer hopping when the two atoms are not
one on top of the other.
It must be noted that, although the first term of Eq. \ref{htb}
maintains electron-hole symmetry, the interlayer coupling
will render the system no longer bipartite, slightly breaking
electron hole symmetry. For the sake of clarity
we have shifted the energy in the
band structures and DOS so that the charge neutrality point
is at zero energy. 
We took $\beta=7 a^{-1}$, and we 
discarded hoppings with a value smaller that $10^{-3}$ $ t_\perp$
to keep the matrices as sparse as possible.
We have verified that changes in $\beta$ do not change the
results qualitatively.
We also explored the effect of second neighbor intralayer hopping,
and we found it does not change qualitatively the
results apart from creating an slightly bigger
electron-hole asymmetry.

The third term in Eq. \ref{htb}
models a perpendicular electric field,
which adds a layer dependent potential $\pm U$
to the atoms in the upper/lower layer,
where $\tau_z$
denotes the layer Pauli matrix
($\tau_z^{ii} = \pm 1$ for the upper/lower layer).
We note that this
form of the electric field does not take into
account the possible screening in the sample,
nor the small corrugation of the layers \cite{PhysRevLett.109.196802}.

\begin{figure}[t!]
 \centering
                \includegraphics[width=\columnwidth]{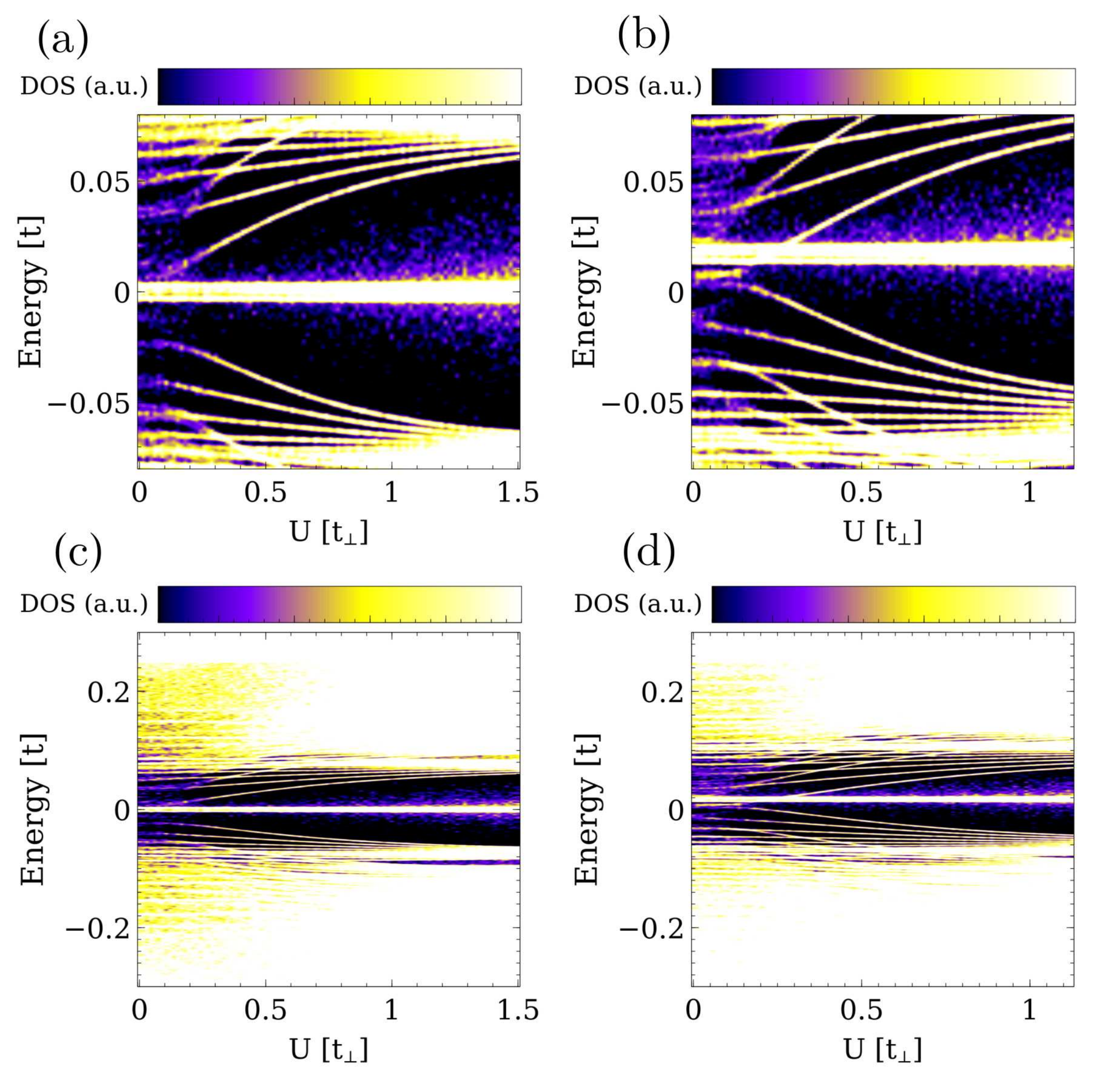}

\caption{
Evolution of twisted bilayer graphene as a function of
the interlayer bias for $m_0=70$ (59644 sites per unit cell)
for two values of the interlayer coupling $t_\perp=0.3t$ (a)
and $t_\perp=0.4t$ (b). 
It is observed that pseudo Landau levels
appear for the two different interlayer couplings,
verifying the scaling argument. 
In particular, spectra for $t_\perp=0.4t$ is equivalent to the
spectra of a bigger unit cell with $t_\perp=0.3t$.
Panels
(c,d) show the same as (a,b) in a bigger energy range,
highlighting that pseudo Landau
levels only appear in the low energy sector, 
in the vicinity of the the charge neutrality point.
The bright states located at zero energy correspond to states located
in the $AA$ region, that remain rather unaffected by the bias.
}
\label{fig1_SM}
\end{figure}

\subsection{Interlayer and twisting scaling}
The angle at which the first flat band shows up is controlled
by the interlayer coupling. In particular, for the real
interlayer coupling of $t_\perp \approx  300$ meV, such angle
corresponds to $\alpha \approx 1^\circ$ \cite{bistritzer2011moire}. 
Changing the value of the
interlayer coupling allows to define at which angle the flat band
appears. In particular, close to the magic angle
the electronic properties
of the system are invariant under the following transformation
$\alpha \rightarrow \alpha/\lambda$ and
$t_{\perp} \rightarrow \lambda t_{\perp}$, 
relation that remains valid for $\alpha \ll1 ^\circ$.
This allows us to reach the tiny
angle regime $\alpha \ll 1^\circ$ with smaller unit cells,
by means of ramping up the interlayer coupling.
It must be noted however that this scaling is valid for
$t_\perp<t$, so that the Dirac cone structure is not destroyed
by the interlayer coupling.
In particular, we can reach the experimental relevant
regime of $\alpha \approx 0.1^\circ$
by taking $m_0=70$ and $t_{\perp} = 0.4 t$.
We have verified that our results are qualitatively unchanged
with different values of $t_\perp$ (see Fig. \ref{fig1_SM}).
Finally, it should be noted that the relevant regime
for pseudo magnetic field requires $U<t_\perp$, and thus scaling
the value of $t_\perp$ allows to observe the pLL
at bigger $U$.

\subsection{Computation of DOS for large systems}

In this section we give computational details on how tight binding calculations
can be performed efficiently for twisted bilayer graphene. We note
that since the unit cells for small angles
are very large, full diagonalization of Hamiltonians
is not generically an option. 
To overcome this
limitation, we calculate the density of states
by using the so called Kernel
polynomial method \cite{RevModPhys.78.275}. This methodology
consists on expanding the density of states
in N Chebyshev polynomials $T_n$, where the coefficients
of the expansion can be calculated by performing
matrix-vector multiplications.

The procedure to calculate the DOS goes as follows.
The first step of the procedure is to
scale the Hamiltonian $H$, so that
new Hamiltonian $\mathcal{H}$
has all its eigenvalues falling in the
interval $(-1,1)$.
For the scaled Hamiltonian
we perform a series expansion of the density of
states as
\begin{equation}
D({\omega}) = \frac{1}{\pi \sqrt{1-\omega^2}}
\left (\mu_n + 2 \sum^N_{n=1} \mu_n T_n (\omega)
\right ).
\label{KPM}
\end{equation}
The coefficients $\mu_n$ determine the expansion of 
the density of states $D(\omega)$, and are expressed as
\begin{equation}
\mu_n = g^N_n \mu^0_n,
\end{equation}
where $\mu^0_n$ are the coefficients calculated as
\begin{equation}
\mu^0_n = 
\sum_i 
\langle i | T_n(\mathcal{H}) | i \rangle
=
\big \langle \langle v | T_n(\mathcal{H}) | v \rangle \big \rangle_v,
\end{equation}
where in the last term we perform the summation using the
stochastic trace method \cite{RevModPhys.78.275}.
The different coefficients are calculated iteratively using
the Chebyshev recursion relation
\begin{eqnarray}
| {w_0} \rangle &=& | {v} \rangle  \\  
|w_1 \rangle &=& \mathcal{H} | w_0 \rangle \\
|w_{n+1} \rangle &=& 2\mathcal{H} |w_n \rangle-
|w_{n-1} \rangle
\end{eqnarray}
so that $ | w_n \rangle = T_n(\mathcal{H}) |v  \rangle$.
The coefficients $\mu^0_n$
are multiplied by  $g_n^N$, defined as:
\begin{equation}
g_n^N =
\frac{(N-n-1)\cos \frac{\pi n}{N+1} + \sin \frac{\pi n}{N+1}
\cot \frac{\pi }{N+1}
}
{N+1},
\end{equation}
which denotes the Jackson Kernel \cite{RevModPhys.78.275}
in order to improve the convergence of the series.

This method
allows us to obtain the density of states in the whole energy
range at once, making it very suitable to
study how the energy spectra evolves in a wide
energy range. The energy resolution $\delta$
of the method is controlled by the number
of terms in the expansion $N$.  In our case,
we performed the Chebyshev expansion with $N=10000$ moments,
which give us a natural broadening of the levels
between $10^{-3}-10^{-4}$t.
Finally, given that we are dealing with a two dimensional periodic
system, this procedure to calculate the density of states
must be performed for the different Bloch Hamiltonians
in the Brillouin zone. We perform this integration
using a Monte Carlo procedure by randomly choosing
$400$ k-points for every different calculation.
The advantage of this procedure is that it allows
us to distinguish the real peaks 
from possible spurious resonances
that would appear in the case of a uniform
k-mesh.

\section{Properties of the flat-band states and their association with pseudo-Landau levels}


In this section we discuss how the emergence of an artificial gauge field due
to interlayer bias in TBLG is associated with the presence of localized states
which we refer to as pseudo-Landau levels (pLL). We examine the properties of
the pLL energies as a function of bias and twist angle, the degree
of localization  of these states as a function of bias and their valley
polarized character. We compare the expected behaviour given by the analytical
treatment with numerical calculations in order to establish a strong connection
between the artificial gauge field and the emergent localized levels observed
in the tight-binding calculation. 

For a real magnetic field $B$, 
the Landau level energies generally follow $E_n = f(n) B^\chi$,
where $f(n)$ is an increasing function of an integer $n$ and $\chi$ is a positive real number. The functional form of $f(n)$ and the value of $\chi$ depend on the details of the low energy dispersion: for Schroedinger electrons
$f(n)\propto n+1/2$ and $\chi=1$, for monolayer graphene\cite{RevModPhys.81.109}
$f(n)\propto \sqrt{n}$ and $\chi=1/2$, for Bernal stacked
bilayer graphene $f(n)\propto \sqrt{n(n-1)}$ and $\chi=1$\cite{mccann2013electronic}
at very low energy, recovering
the monolayer result at higher energies.\cite{mccann2013electronic}
Independently of the details
of the low energy dispersion, for
$n>0$,  electron-like Landau level energies always follow $\partial E_n/\partial B>0$.

From our analytical calculation, we find that the artificial gauge fields
$g_{x,y} (\br)$, given by Eqs. 13-14 in the main text, are written in terms of
$\gamma_{x,y}(\br)$ and $A_{x,y}(\br)$ (given in Eq. 12 in the main text),
which in turn are written in terms of $a_{x,y}(\br)$, $v_{x,y}(\br)$ (provided
in the SM, Eqs. 2 and 4, respectively) and $\partial U(\br)$. One can show that
for small values of $U(\br)$ the functions $a_{x,y}(\br)$ and $v_{x,y}(\br)$
are finite and have only small corrections proportional to $U(\br)$, which can
in principle be neglected. The main dependence of the gauge field on the
external bias is due to the explicit linear dependence of $A_{x,y}(\br)$ on
$\partial U(\br)$ given by Eq. 12. 
The effective model given in Eq. 2 describes TBLG under an external bias by the
introduction of a modulated inter-layer bias $U(\br)$ in a model with only two
effective degrees of freedom. This inter-layer bias has a characteristic
wave-vector which is proportional to the inverse Moire length, $k_M=2\pi/L_M$,
and can be minimally written as $U(r) \sim U \cos(k_M r)$, where $U$ characterizes
its magnitude. As the gauge field depends linearly on $\partial U(\br)$, we
find $g_{x,y} (\br) \sim U k_M$. Remembering that the Moire length $L_M=
a/[2\sin(\alpha/2)]$, where $a$ is the microscopic lattice constant and
$\alpha$ the twist angle, for small angles we can write: $g_{x,y} (\br) \sim
U \alpha$, what explicitly shows that the gauge field is proportional to the
magnitude of the bias and to the twist angle.

From this discussion, there are three properties of the flat band energies and
associated levels which can be verified within the tight-binding calculations
to establish that the states associated with these levels originate
from an
artificial gauge field induced by the electric bias and behave as pseudo-Landau
levels:

\begin{itemize}

\item One signature of the pLL nature of the states in the flat bands is the increase of their energies with the magnitude of the artificial gauge field. In analogy to the standard case of electrons under a real magnetic field, we expect the energies of the pLL to increase with the magnitude of the artificial gauge-field $\partial E/\partial g>0$, where $g = \sqrt{g_x^2+g_y^2}$. As discussed above, $g \sim U$, therefore we expect $\partial E/\partial U>0$.

\item A second signature is the behaviour of the characteristic localization
	length of these states as a function of the magnitude of the artificial
		gauge field. For standard Landau levels, the localization
		length is determined
by the magnetic length $l_B\propto 1/\sqrt{B}$, so that the states become more localized as $B$ increases.
For biased tiny angle twisted bilayer graphene, the bias dependent gauge field $g \sim U$, so that it is expected that the pLL become more localized as the interlayer bias $U$ increases.

\item Last but not least, the analytic derivation of the emergent
gauge field relied on a single valley Hamiltonian, so that a
necessary condition for the analytic and numerical results to be compatible
is that the Landau levels are valley polarized.

\end{itemize}

In the following subsections we verify these three properties within the tight-binding calculation.

\begin{figure}[t!]
 \centering
                \includegraphics[width=\columnwidth]{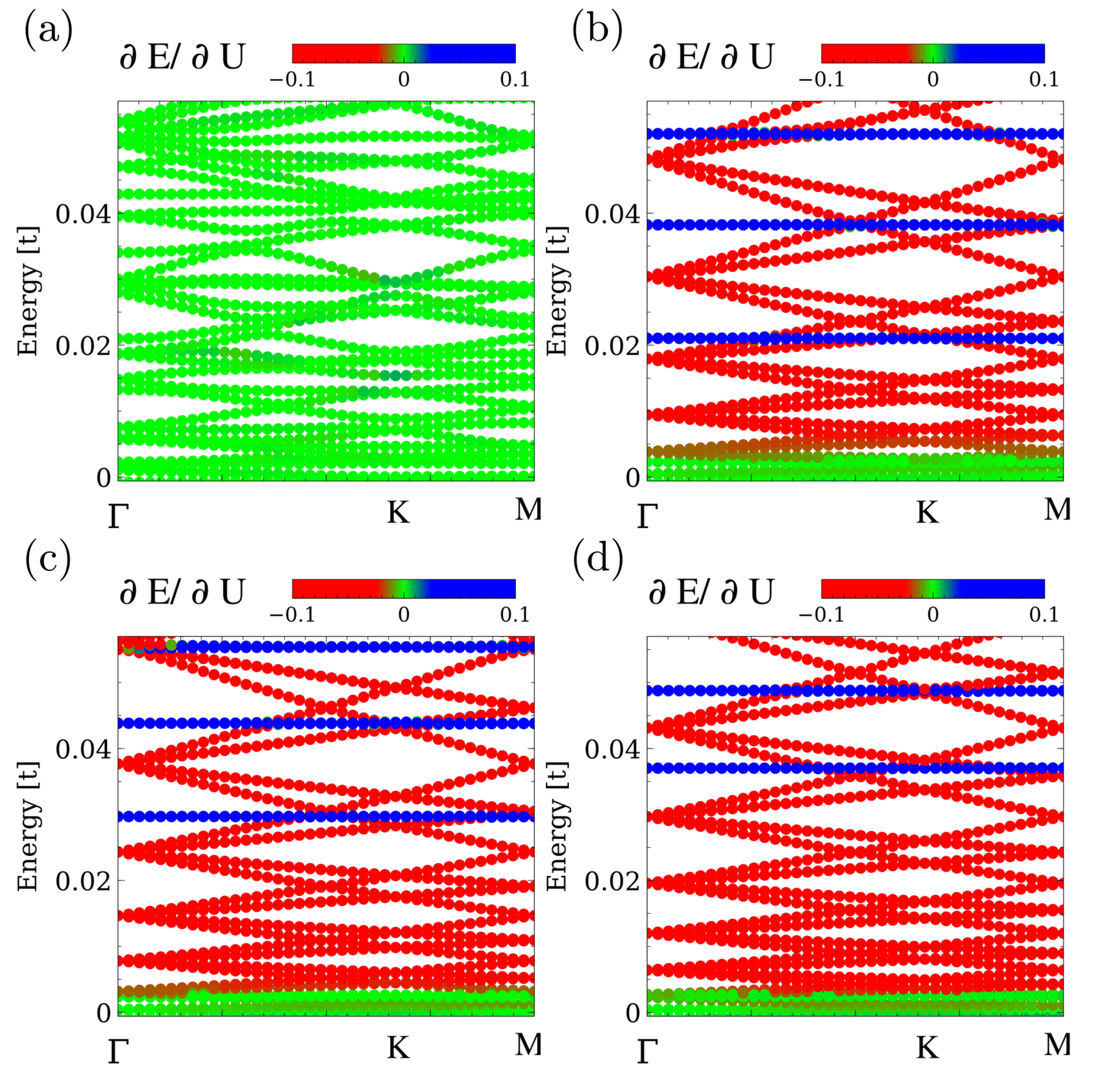}

\caption{Comparison of the band structure of tiny angle twisted
	bilayer graphene for (a) $U=0$, (b) $U=0.5t_{\perp}$, (c) $U=0.625t_{\perp}$ and (d)
	$U=0.75t_{\perp}$.
	The color of the energy levels represents the expectation value
	of $\partial E/\partial U$. It is observed that for $U=0$
	the states do not evolve with $U$. In contrast,
	for $U\neq0$, the pLL increase their energy with $U$ (blue),
	the $AA$ modes remain unmodified (green) and the helical
	network states flow towards charge neutrality (red).
	This highlights that for $U=0$ no flat
	bands associated with pseudo Landau levels
	are present.
	We took for these calculations $m_0=75$ and $t_\perp=0.4t$.
}
\label{fig2_SM}
\end{figure}

\subsection{Energy evolution with interlayer bias and twist angle}

The first signature that can be benchmarked is the bias and twist angle
dependence of the energy levels. In order to do that, it is convenient to use
the Hellmann-Feynman theorem to the
Hamiltonian in Eq. \ref{htb} to compute the energy dependence on the bias
of a certain
state $\Psi$ as
\begin{equation}
\frac{\partial E}{\partial U} = \langle \Psi | \tau_z | \Psi \rangle.
	\label{flow}
\end{equation} 
Here $\tau_z$ is a Pauli matrix related to the layer degree of freedom. This
quantity characterizes
the evolution of the energy of a state when the bias is increased,
in particular whether the energy will increase 
($\partial E/\partial U>0$),
decrease ($\partial E/\partial U<0$),
or stay the same ($\partial E/\partial U=0$).
In particular, for AA modes it is expected
$\partial E/\partial U=0$, given that they are bonding-antibonding
combinations of the two layers \cite{PhysRevB.90.155415}.
In constrast, for pseudo Landau levels
we expect $\partial E/\partial U>0$ given that $g\sim U$, so an increasing a gauge field will increase
the energy of the associated Landau levels. Finally, since the interlayer
bias will increase the density of states in the AA regions for $E=0$, there
must be a flow of states towards the charge
neutrality point \cite{PhysRevB.88.121408}, and as we will
see shortly these states are associated with helical network modes.

We now move on to compute the expectation value
presented in Eq. \ref{flow} with the numerically
computed wavefunctions obtained with the real space tight-binding model.
In particular, we will focus on the case $U=0$, which does not show
bias induced pLLs and the case $U\neq0$, which displays flat band
pLLs.
In the case without interlayer bias $U=0$, the low energy
states of the system are located mainly at the AA regions. 
As mentioned before, these states
are to first order insensitive to the interlayer bias, due to their
bonding-antibonding nature between the two layers,\cite{PhysRevB.90.155415}
and thus remain nearly unmodified by the bias.

The most interesting scenario concerns the case of $U\neq0$ when
the system is in the pseudo Landau level regime. As anticipated
above, in this case the three different sets of states (pseudo Landau levels,
AA modes and helical network states) show very distinct
behavior. In particular, the value of $\partial E/\partial U$
allows to identify the physical nature of each state: AA modes
remain unmodified by the interlayer bias
(green in Fig. \ref{fig2_SM}b),
pseudo Landau levels increase their energy with $U$ due to
the increase of the effective gauge field
(blue in Fig. \ref{fig2_SM}b),
and helical network states flow towards
the charge neutrality point
(red in Fig. \ref{fig2_SM}b).
In this way, Eq. \ref{flow} allows to distinguish pseudo Landau levels
from other modes (AA modes or helical network states),
due to the distinct dependence of the respective energy levels on the bias.

\begin{figure}[t!]
 \centering
                \includegraphics[width=\columnwidth]{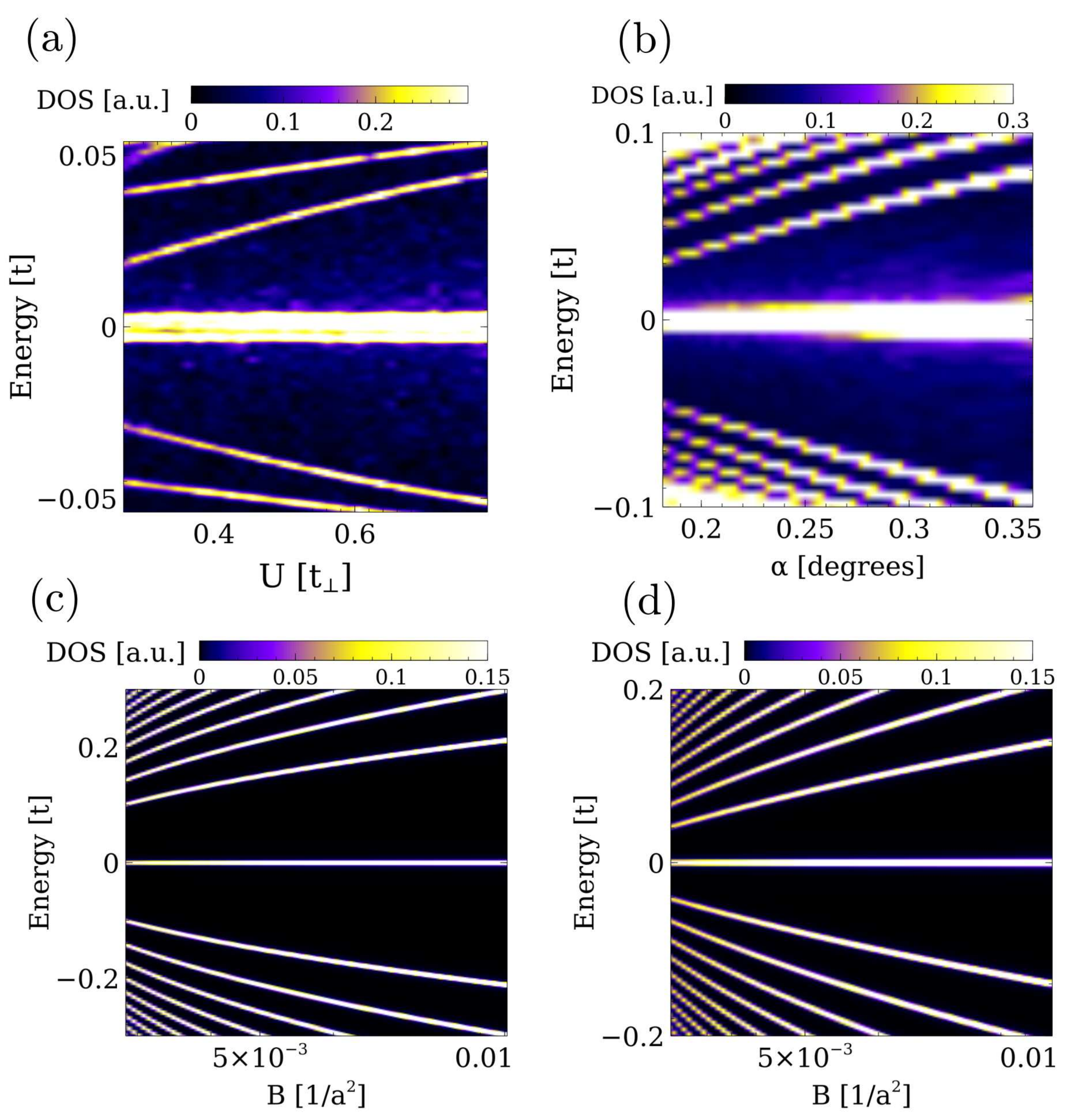}

\caption{ Evolution of the spectra of tiny angle
	 twisted bilayer graphene with interlayer bias $U$ (a)
	and twist angle $\alpha$ (b) for $U\neq0$. For comparison, we show in panels (c) and (d)
	the Landau level spectra of monolayer (c) and bilayer AB (d)
	graphene under a true magnetic field $B$, showing a similar
	evolution as the pseudo Landau levels of panels (a) and (b) for which the
	artificial gauge field is controlled by $\alpha$ and $U$.
	Panels (a) and (b) contain the same information as panels (b) and (d) of Fig. \ref{fig2}bd.
}
\label{fig5_SM}
\end{figure}

Another property which we can benchmark is the dependence of the energies of
the states in the flat bands on the twist angle. First, we note that in the
following discussion we will be all the time in the tiny angle limit $\alpha
\ll 1$, so that the mapping to a gauge field remains valid. As discussed above,
the bias controlled gauge field $g$ depends explicitly on the spatial
derivative of the modulated bias, therefore it is proportional to the inverse
Moire length $L_M$. In the tiny angle regime $L_M \sim 1/\alpha$, so ultimately
the magnitude of the artificial gauge field is proportional to the twist angle.

To verify this prediction obtained analytically with the numerical results, we show in Figs. \ref{fig5_SM} (a) and (b)
the evolution of the pseudo Landau levels with the magnitude of the bias and with the twist angle $\alpha$
between the two layers, respectively. It is observed that the energies
of the pseudo Landau levels increase linearly in both cases.
We note that the zero pseudo-Landau level cannot be clearly distinguished due to the mixing
with the AA modes.

For comparison, we
show in Figs. \ref{fig5_SM} (c) and (d) the Landau level spectra of monolayer
graphene and AB bilayer graphene, respectively,   in the presence of a real
magnetic field $B$. 
In the case of a real magnetic field $B$, it is observed that the
evolution of the Landau levels is slightly different for monolayer
and bilayer, yet their dependence with magnetic field $B$ strongly
resembles the evolution of the pseudo Landau levels of twisted
bilayer graphene with $U$ and $\alpha$. Figs. \ref{fig5_SM} (c) and (d)
are obtained using a first neighbor tight
binding Hamiltonian, including the magnetic field $B$
by a Peierls substitution in a monolayer/bilayer ribbon
with 600 sites per unit cell, and computing
the density of states in the bulk of the ribbon. The magnetic field
is measured in units of the carbon-carbon distance $a$.

%

\subsection{Bias dependent localization}

\begin{figure}[t!]
 \centering
                \includegraphics[width=\columnwidth]{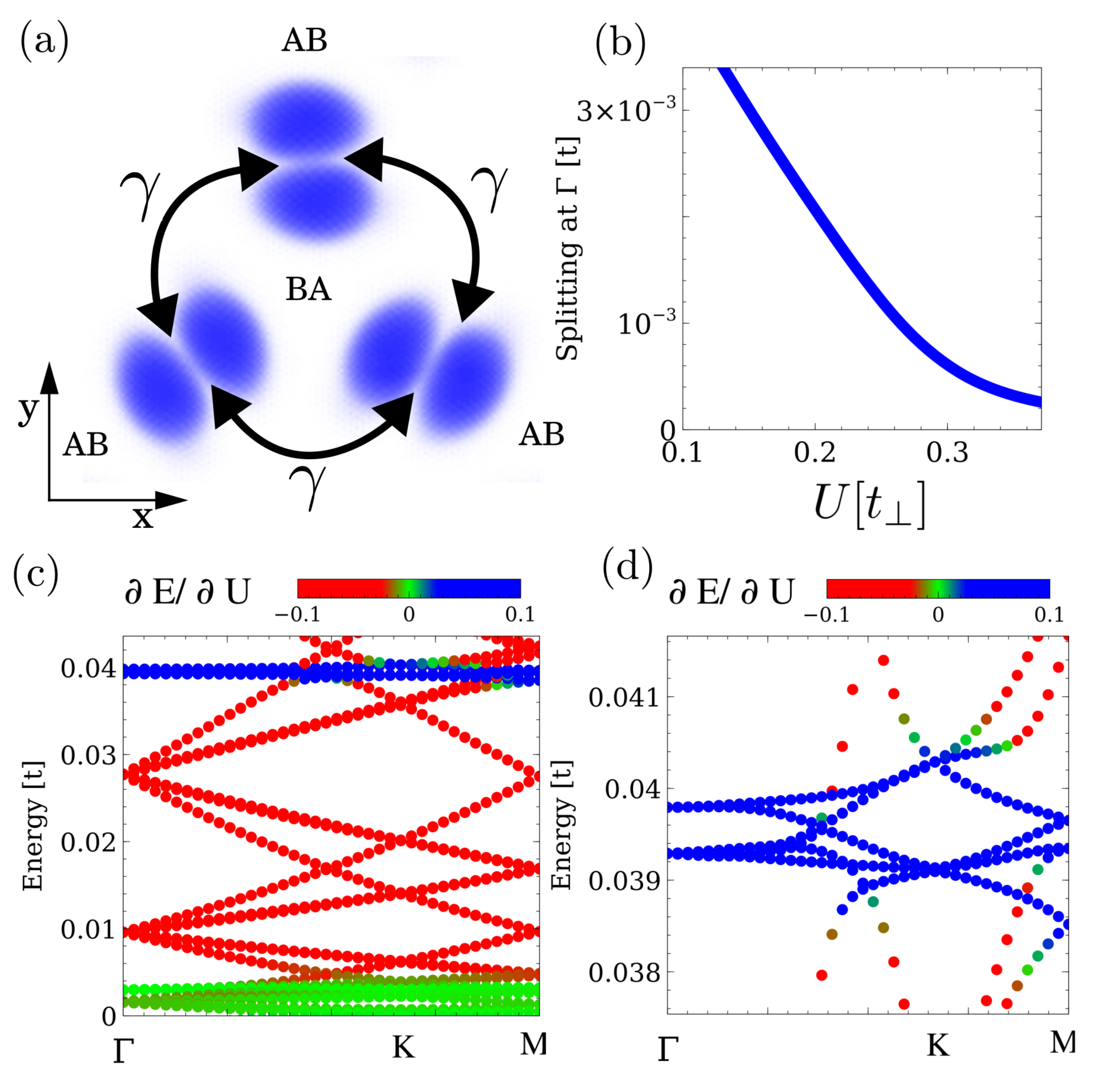}

\caption{ 
	Sketch of the hoppings between pseudo Landau levels (a),
	located in the emergent Kagome lattice
	whose lattice constant is the Moire length $L_M$. Panel (b)
	shows the evolution of the splitting at the
	$\Gamma$ point of the first pseudo
	Landau levels highlighted in the 
	band structure of panels (c) and (d).
	The evolution of the splitting in (b) highlights that
	as the bias $U$ is increased, the pseudo Landau levels become
	more localized as expected from an artificial gauge field.
	Panel (d) is a zoom in the pseudo Landau levels of panel (c).
	The parameters used are $m_0=75$ and $t_{\perp}=0.3t$.
}
\label{fig3_SM}
\end{figure}

We now move on to consider the localization of the states as a function
of interlayer bias. A simple way to characterize the localization
of the pseudo Landau levels is by computing the bandwidth of their associated
bands. First, it is worthy to first remark that the first pseudo Landau levels
form an emergent Kagome lattice, with pseudo Landau levels
located at the interfaces between AB/BA regions. 
For these localized modes, one can consider 
an effective tight binding Hamiltonian $H_W$ defined
for the Wannier functions of the pseudo Landau levels
$
	H_W = \sum_{ij}\gamma_{ij} \Psi^\dagger_i \Psi_j
	$
with Wannier orbitals located in the different
sites of the emergent Kagome lattice.
$\Psi^\dagger_i$ ($\Psi_i$) are creation (annihilation) operators
for the pseudo Landau levels in Kagome site $i$, and
$\gamma_{ij}$ their effective hopping
(see Fig. \ref{fig3_SM}a). The
bandwidth of the pseudo Landau level bands is proportional
to the hoppings $\gamma_{ij}$,
and will decrease to zero as the states become more localized.
In this way, the bandwidth of the pLL bands reflects the spatial extension
of the states. Following the previous argument, we plot in Fig.
\ref{fig3_SM}b the evolution of the splitting of the
pseudo Landau level bands (blue bands in Fig. \ref{fig3_SM}c, zoomed
in Fig. \ref{fig3_SM}d)
at the $\Gamma$ point as a function the interlayer bias. It is observed
that the splitting goes to zero as the bias $U$ is increased,
signaling that the states become more localized. This behavior
is the one expected for pLL arising from a gauge field proportional
to the interlayer bias $U$.

From the computational point of view, the pLL can be systematically targeted by
retaining the lowest energy states with $dE/dU>0$ as discussed in the
previous section (Fig. \ref{fig2_SM}b).
We finally note that although we used the splitting
at the $\Gamma$ point as an estimate of the hopping parameter, the maximum
bandwidth of the pLL bands may be located at other points
of the Brillouin zone (see Fig. \ref{fig3_SM}d).

\begin{figure}[t!]
 \centering
                \includegraphics[width=\columnwidth]{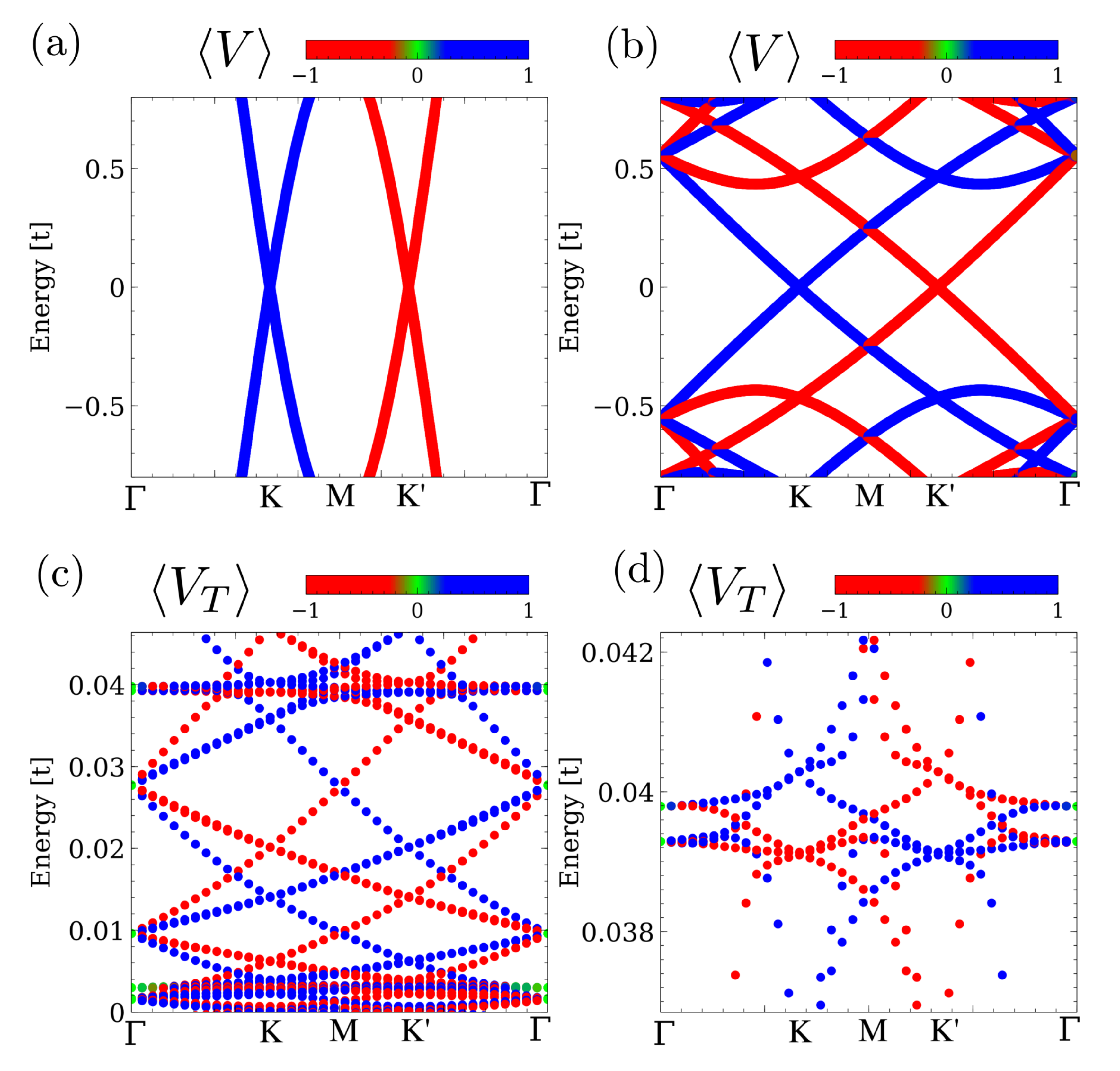}

\caption{
	Expectation value of the valley operator
	for the monolayer graphene with a minimal unit cell (a),  and with a 7x7
	unit cell (b).
	Expectation value of the valley operator for the
	upper layer of a tiny angle twisted bilayer graphene
	with interlayer bias showing the valley
	polarization of both helical network states (c)
	and pseudo Landau level bands (d).
	The parameters used in (cd) are $m_0=75$ and $t_{\perp}=0.3t$.
}
\label{fig4_SM}
\end{figure}

\subsection{Valley polarization}
 In this section we show that the pseudo Landau level flat
 bands generated by the interlayer bias are valley polarized. This is an important feature which supports the use of an effective low energy model with a single valley.

 We first note that we are working numerically with a real space
 tight binding Hamiltonian, so the concept of
 valley is an emergent aspect of the reciprocal space band structure. In principle, it would seem
 that there is no good representation of the valley operator in a real space
 tight binding basis. Nevertheless, it is possible to compute the expectation value of a valley operator
 for a certain state, by choosing a minimal tight binding model whose
 eigenvalues are $\epsilon>0$ around the $K$ point and $\epsilon<0$
 around the $K'$ point, or in other words, a valley dependent chemical
 potential. A tight binding model with these properties is given
 by a sublattice dependent Haldane coupling\cite{PhysRevLett.120.086603} of the form
 \begin{equation}
	 \hat V =\frac{i}{3\sqrt{3}} \sum_{\langle\langle ij \rangle\rangle}
	 \eta_{ij} \sigma_z^{ij} c^\dagger_i c_j,
	 \label{antihaldane}
 \end{equation}
 where $\langle\langle ij \rangle\rangle$ denotes second neighbor
 sites, $\eta_{ij}=\pm 1$ for clockwise or anticlockwise hopping and
 $\sigma^{ij}_z$ is a Pauli matrix associated with the sublattice degree of freedom. 
 The previous real space operator has eigenvalues $\pm 1$ for states close to the $K$ and $K'$ points,\cite{PhysRevLett.120.086603}
 which implies that it can be used to evaluate the expectation value of the valley degree
 of freedom for an arbitrary state.
 With this operator at hand, the valley expectation value $\langle V\rangle $ can be computed as
 \begin{equation}
	  \langle V\rangle  = \langle \Psi | \hat V | \Psi \rangle.
 \end{equation}
 
 The previous methodology yields a procedure to compute the valley expectation
 value in a real space tight binding model. As particular examples,
 we show in Fig. \ref{fig4_SM}ab the bandstructure together with
 the valley expectation value for the minimal graphene supercell
 (Fig. \ref{fig4_SM}a) and for a 7x7 graphene supercell (Fig. \ref{fig4_SM}b).
 In particular, as shown in the bandstructure of the 7x7 graphene
 supercell in Fig. \ref{fig4_SM}b, the valley operator allows us to identify
 the original valley flavor of a state even when the bandstructure
 is heavily folded in case the unit cell includes many carbon atoms. 
 This procedure becomes specially useful
 to identify the valley flavor a state in twisted bilayer graphene
 starting from a real space tight binding model.

We now move on to compute the valley expectation value for
biased twisted bilayer graphene. In this case, we use
as valley projector $\hat V_T$ an operator analogous to Eq. \ref{antihaldane}
but only considering sites in the top layer $T$
 \begin{equation}
	 \hat V_T =
	 \frac{i}{3\sqrt{3}}
	 \sum_{\langle\langle ij \rangle\rangle\text{ and } ij \in T}
         \eta_{ij} \sigma_z^{ij} c^\dagger_i c_j.
 \end{equation}

 With the previous operator we compute the upper layer
 valley expectation value
 of the states computed numerically as
 \begin{equation}
	 \langle V_T \rangle = \langle \Psi | \hat V_T | \Psi \rangle.
 \end{equation}
 We now apply this operator for the
 biased tiny angle twisted bilayer graphene, as shown in Fig. \ref{fig4_SM}c
 (with a zoom into the pseudo Landau levels in Fig. \ref{fig4_SM}d).
 It is clearly observed that the states show a nearly perfect valley
 polarization. 
 Therefore, this calculation supports the analytical model used, which takes
 into account a single valley and thus does not consider intervalley
 mixing effects.

\bibliographystyle{apsrev4-1}
\bibliography{paper}{}

\begin{thebibliography}{58}%
\makeatletter
\providecommand \@ifxundefined [1]{%
 \@ifx{#1\undefined}
}%
\providecommand \@ifnum [1]{%
 \ifnum #1\expandafter \@firstoftwo
 \else \expandafter \@secondoftwo
 \fi
}%
\providecommand \@ifx [1]{%
 \ifx #1\expandafter \@firstoftwo
 \else \expandafter \@secondoftwo
 \fi
}%
\providecommand \natexlab [1]{#1}%
\providecommand \enquote  [1]{``#1''}%
\providecommand \bibnamefont  [1]{#1}%
\providecommand \bibfnamefont [1]{#1}%
\providecommand \citenamefont [1]{#1}%
\providecommand \href@noop [0]{\@secondoftwo}%
\providecommand \href [0]{\begingroup \@sanitize@url \@href}%
\providecommand \@href[1]{\@@startlink{#1}\@@href}%
\providecommand \@@href[1]{\endgroup#1\@@endlink}%
\providecommand \@sanitize@url [0]{\catcode `\\12\catcode `\$12\catcode
  `\&12\catcode `\#12\catcode `\^12\catcode `\_12\catcode `\%12\relax}%
\providecommand \@@startlink[1]{}%
\providecommand \@@endlink[0]{}%
\providecommand \url  [0]{\begingroup\@sanitize@url \@url }%
\providecommand \@url [1]{\endgroup\@href {#1}{\urlprefix }}%
\providecommand \urlprefix  [0]{URL }%
\providecommand \Eprint [0]{\href }%
\providecommand \doibase [0]{http://dx.doi.org/}%
\providecommand \selectlanguage [0]{\@gobble}%
\providecommand \bibinfo  [0]{\@secondoftwo}%
\providecommand \bibfield  [0]{\@secondoftwo}%
\providecommand \translation [1]{[#1]}%
\providecommand \BibitemOpen [0]{}%
\providecommand \bibitemStop [0]{}%
\providecommand \bibitemNoStop [0]{.\EOS\space}%
\providecommand \EOS [0]{\spacefactor3000\relax}%
\providecommand \BibitemShut  [1]{\csname bibitem#1\endcsname}%
\let\auto@bib@innerbib\@empty
\bibitem [{\citenamefont {Castro~Neto}\ \emph {et~al.}(2009)\citenamefont
  {Castro~Neto}, \citenamefont {Guinea}, \citenamefont {Peres}, \citenamefont
  {Novoselov},\ and\ \citenamefont {Geim}}]{RevModPhys.81.109}%
  \BibitemOpen
  \bibfield  {author} {\bibinfo {author} {\bibfnamefont {A.~H.}\ \bibnamefont
  {Castro~Neto}}, \bibinfo {author} {\bibfnamefont {F.}~\bibnamefont {Guinea}},
  \bibinfo {author} {\bibfnamefont {N.~M.~R.}\ \bibnamefont {Peres}}, \bibinfo
  {author} {\bibfnamefont {K.~S.}\ \bibnamefont {Novoselov}}, \ and\ \bibinfo
  {author} {\bibfnamefont {A.~K.}\ \bibnamefont {Geim}},\ }\href {\doibase
  10.1103/RevModPhys.81.109} {\bibfield  {journal} {\bibinfo  {journal} {Rev.
  Mod. Phys.}\ }\textbf {\bibinfo {volume} {81}},\ \bibinfo {pages} {109}
  (\bibinfo {year} {2009})}\BibitemShut {NoStop}%
\bibitem [{\citenamefont {Novoselov}\ \emph {et~al.}(2007)\citenamefont
  {Novoselov}, \citenamefont {Jiang}, \citenamefont {Zhang}, \citenamefont
  {Morozov}, \citenamefont {Stormer}, \citenamefont {Zeitler}, \citenamefont
  {Maan}, \citenamefont {Boebinger}, \citenamefont {Kim},\ and\ \citenamefont
  {Geim}}]{novoselov2007room}%
  \BibitemOpen
  \bibfield  {author} {\bibinfo {author} {\bibfnamefont {K.~S.}\ \bibnamefont
  {Novoselov}}, \bibinfo {author} {\bibfnamefont {Z.}~\bibnamefont {Jiang}},
  \bibinfo {author} {\bibfnamefont {Y.}~\bibnamefont {Zhang}}, \bibinfo
  {author} {\bibfnamefont {S.~V.}\ \bibnamefont {Morozov}}, \bibinfo {author}
  {\bibfnamefont {H.~L.}\ \bibnamefont {Stormer}}, \bibinfo {author}
  {\bibfnamefont {U.}~\bibnamefont {Zeitler}}, \bibinfo {author} {\bibfnamefont
  {J.~C.}\ \bibnamefont {Maan}}, \bibinfo {author} {\bibfnamefont {G.~S.}\
  \bibnamefont {Boebinger}}, \bibinfo {author} {\bibfnamefont {P.}~\bibnamefont
  {Kim}}, \ and\ \bibinfo {author} {\bibfnamefont {A.~K.}\ \bibnamefont
  {Geim}},\ }\href {\doibase 10.1126/science.1137201} {\bibfield  {journal}
  {\bibinfo  {journal} {Science}\ }\textbf {\bibinfo {volume} {315}},\ \bibinfo
  {pages} {1379} (\bibinfo {year} {2007})}\BibitemShut {NoStop}%
\bibitem [{\citenamefont {Zhang}\ \emph {et~al.}(2005)\citenamefont {Zhang},
  \citenamefont {Tan}, \citenamefont {Stormer},\ and\ \citenamefont
  {Kim}}]{zhang2005experimental}%
  \BibitemOpen
  \bibfield  {author} {\bibinfo {author} {\bibfnamefont {Y.}~\bibnamefont
  {Zhang}}, \bibinfo {author} {\bibfnamefont {Y.-W.}\ \bibnamefont {Tan}},
  \bibinfo {author} {\bibfnamefont {H.~L.}\ \bibnamefont {Stormer}}, \ and\
  \bibinfo {author} {\bibfnamefont {P.}~\bibnamefont {Kim}},\ }\href {\doibase
  10.1038/nature04235} {\bibfield  {journal} {\bibinfo  {journal} {Nature}\
  }\textbf {\bibinfo {volume} {438}},\ \bibinfo {pages} {201} (\bibinfo {year}
  {2005})}\BibitemShut {NoStop}%
\bibitem [{\citenamefont {Yang}(2007)}]{yang2007spontaneous}%
  \BibitemOpen
  \bibfield  {author} {\bibinfo {author} {\bibfnamefont {K.}~\bibnamefont
  {Yang}},\ }\href@noop {} {\bibfield  {journal} {\bibinfo  {journal} {Solid
  State Communications}\ }\textbf {\bibinfo {volume} {143}},\ \bibinfo {pages}
  {27} (\bibinfo {year} {2007})}\BibitemShut {NoStop}%
\bibitem [{\citenamefont {Bolotin}\ \emph {et~al.}(2009)\citenamefont
  {Bolotin}, \citenamefont {Ghahari}, \citenamefont {Shulman}, \citenamefont
  {Stormer},\ and\ \citenamefont {Kim}}]{bolotin2009observation}%
  \BibitemOpen
  \bibfield  {author} {\bibinfo {author} {\bibfnamefont {K.~I.}\ \bibnamefont
  {Bolotin}}, \bibinfo {author} {\bibfnamefont {F.}~\bibnamefont {Ghahari}},
  \bibinfo {author} {\bibfnamefont {M.~D.}\ \bibnamefont {Shulman}}, \bibinfo
  {author} {\bibfnamefont {H.~L.}\ \bibnamefont {Stormer}}, \ and\ \bibinfo
  {author} {\bibfnamefont {P.}~\bibnamefont {Kim}},\ }\href {\doibase
  10.1038/nature08582} {\bibfield  {journal} {\bibinfo  {journal} {Nature}\
  }\textbf {\bibinfo {volume} {462}},\ \bibinfo {pages} {196} (\bibinfo {year}
  {2009})}\BibitemShut {NoStop}%
\bibitem [{\citenamefont {Young}\ \emph {et~al.}(2013)\citenamefont {Young},
  \citenamefont {Sanchez-Yamagishi}, \citenamefont {Hunt}, \citenamefont
  {Choi}, \citenamefont {Watanabe}, \citenamefont {Taniguchi}, \citenamefont
  {Ashoori},\ and\ \citenamefont {Jarillo-Herrero}}]{young2014tunable}%
  \BibitemOpen
  \bibfield  {author} {\bibinfo {author} {\bibfnamefont {A.~F.}\ \bibnamefont
  {Young}}, \bibinfo {author} {\bibfnamefont {J.~D.}\ \bibnamefont
  {Sanchez-Yamagishi}}, \bibinfo {author} {\bibfnamefont {B.}~\bibnamefont
  {Hunt}}, \bibinfo {author} {\bibfnamefont {S.~H.}\ \bibnamefont {Choi}},
  \bibinfo {author} {\bibfnamefont {K.}~\bibnamefont {Watanabe}}, \bibinfo
  {author} {\bibfnamefont {T.}~\bibnamefont {Taniguchi}}, \bibinfo {author}
  {\bibfnamefont {R.~C.}\ \bibnamefont {Ashoori}}, \ and\ \bibinfo {author}
  {\bibfnamefont {P.}~\bibnamefont {Jarillo-Herrero}},\ }\href {\doibase
  10.1038/nature12800} {\bibfield  {journal} {\bibinfo  {journal} {Nature}\
  }\textbf {\bibinfo {volume} {505}},\ \bibinfo {pages} {528} (\bibinfo {year}
  {2013})}\BibitemShut {NoStop}%
\bibitem [{\citenamefont {Young}\ \emph {et~al.}(2012)\citenamefont {Young},
  \citenamefont {Dean}, \citenamefont {Wang}, \citenamefont {Ren},
  \citenamefont {Cadden-Zimansky}, \citenamefont {Watanabe}, \citenamefont
  {Taniguchi}, \citenamefont {Hone}, \citenamefont {Shepard},\ and\
  \citenamefont {Kim}}]{young2012spin}%
  \BibitemOpen
  \bibfield  {author} {\bibinfo {author} {\bibfnamefont {A.~F.}\ \bibnamefont
  {Young}}, \bibinfo {author} {\bibfnamefont {C.~R.}\ \bibnamefont {Dean}},
  \bibinfo {author} {\bibfnamefont {L.}~\bibnamefont {Wang}}, \bibinfo {author}
  {\bibfnamefont {H.}~\bibnamefont {Ren}}, \bibinfo {author} {\bibfnamefont
  {P.}~\bibnamefont {Cadden-Zimansky}}, \bibinfo {author} {\bibfnamefont
  {K.}~\bibnamefont {Watanabe}}, \bibinfo {author} {\bibfnamefont
  {T.}~\bibnamefont {Taniguchi}}, \bibinfo {author} {\bibfnamefont
  {J.}~\bibnamefont {Hone}}, \bibinfo {author} {\bibfnamefont {K.~L.}\
  \bibnamefont {Shepard}}, \ and\ \bibinfo {author} {\bibfnamefont
  {P.}~\bibnamefont {Kim}},\ }\href {\doibase 10.1038/nphys2307} {\bibfield
  {journal} {\bibinfo  {journal} {Nature Physics}\ }\textbf {\bibinfo {volume}
  {8}},\ \bibinfo {pages} {550} (\bibinfo {year} {2012})}\BibitemShut {NoStop}%
\bibitem [{\citenamefont {Takei}\ \emph {et~al.}(2016)\citenamefont {Takei},
  \citenamefont {Yacoby}, \citenamefont {Halperin},\ and\ \citenamefont
  {Tserkovnyak}}]{PhysRevLett.116.216801}%
  \BibitemOpen
  \bibfield  {author} {\bibinfo {author} {\bibfnamefont {S.}~\bibnamefont
  {Takei}}, \bibinfo {author} {\bibfnamefont {A.}~\bibnamefont {Yacoby}},
  \bibinfo {author} {\bibfnamefont {B.~I.}\ \bibnamefont {Halperin}}, \ and\
  \bibinfo {author} {\bibfnamefont {Y.}~\bibnamefont {Tserkovnyak}},\ }\href
  {\doibase 10.1103/PhysRevLett.116.216801} {\bibfield  {journal} {\bibinfo
  {journal} {Phys. Rev. Lett.}\ }\textbf {\bibinfo {volume} {116}},\ \bibinfo
  {pages} {216801} (\bibinfo {year} {2016})}\BibitemShut {NoStop}%
\bibitem [{\citenamefont {Stepanov}\ \emph {et~al.}(2018)\citenamefont
  {Stepanov}, \citenamefont {Che}, \citenamefont {Shcherbakov}, \citenamefont
  {Yang}, \citenamefont {Chen}, \citenamefont {Thilahar}, \citenamefont
  {Voigt}, \citenamefont {Bockrath}, \citenamefont {Smirnov}, \citenamefont
  {Watanabe}, \citenamefont {Taniguchi}, \citenamefont {Lake}, \citenamefont
  {Barlas}, \citenamefont {MacDonald},\ and\ \citenamefont
  {Lau}}]{stepanov2018long}%
  \BibitemOpen
  \bibfield  {author} {\bibinfo {author} {\bibfnamefont {P.}~\bibnamefont
  {Stepanov}}, \bibinfo {author} {\bibfnamefont {S.}~\bibnamefont {Che}},
  \bibinfo {author} {\bibfnamefont {D.}~\bibnamefont {Shcherbakov}}, \bibinfo
  {author} {\bibfnamefont {J.}~\bibnamefont {Yang}}, \bibinfo {author}
  {\bibfnamefont {R.}~\bibnamefont {Chen}}, \bibinfo {author} {\bibfnamefont
  {K.}~\bibnamefont {Thilahar}}, \bibinfo {author} {\bibfnamefont
  {G.}~\bibnamefont {Voigt}}, \bibinfo {author} {\bibfnamefont {M.~W.}\
  \bibnamefont {Bockrath}}, \bibinfo {author} {\bibfnamefont {D.}~\bibnamefont
  {Smirnov}}, \bibinfo {author} {\bibfnamefont {K.}~\bibnamefont {Watanabe}},
  \bibinfo {author} {\bibfnamefont {T.}~\bibnamefont {Taniguchi}}, \bibinfo
  {author} {\bibfnamefont {R.~K.}\ \bibnamefont {Lake}}, \bibinfo {author}
  {\bibfnamefont {Y.}~\bibnamefont {Barlas}}, \bibinfo {author} {\bibfnamefont
  {A.~H.}\ \bibnamefont {MacDonald}}, \ and\ \bibinfo {author} {\bibfnamefont
  {C.~N.}\ \bibnamefont {Lau}},\ }\href {\doibase 10.1038/s41567-018-0161-5}
  {\bibfield  {journal} {\bibinfo  {journal} {Nature Physics}\ }\textbf
  {\bibinfo {volume} {14}},\ \bibinfo {pages} {907} (\bibinfo {year}
  {2018})}\BibitemShut {NoStop}%
\bibitem [{\citenamefont {Levy}\ \emph {et~al.}(2010)\citenamefont {Levy},
  \citenamefont {Burke}, \citenamefont {Meaker}, \citenamefont {Panlasigui},
  \citenamefont {Zettl}, \citenamefont {Guinea}, \citenamefont {Neto},\ and\
  \citenamefont {Crommie}}]{Levy2010}%
  \BibitemOpen
  \bibfield  {author} {\bibinfo {author} {\bibfnamefont {N.}~\bibnamefont
  {Levy}}, \bibinfo {author} {\bibfnamefont {S.~A.}\ \bibnamefont {Burke}},
  \bibinfo {author} {\bibfnamefont {K.~L.}\ \bibnamefont {Meaker}}, \bibinfo
  {author} {\bibfnamefont {M.}~\bibnamefont {Panlasigui}}, \bibinfo {author}
  {\bibfnamefont {A.}~\bibnamefont {Zettl}}, \bibinfo {author} {\bibfnamefont
  {F.}~\bibnamefont {Guinea}}, \bibinfo {author} {\bibfnamefont {A.~H.~C.}\
  \bibnamefont {Neto}}, \ and\ \bibinfo {author} {\bibfnamefont {M.~F.}\
  \bibnamefont {Crommie}},\ }\href {\doibase 10.1126/science.1191700}
  {\bibfield  {journal} {\bibinfo  {journal} {Science}\ }\textbf {\bibinfo
  {volume} {329}},\ \bibinfo {pages} {544} (\bibinfo {year}
  {2010})}\BibitemShut {NoStop}%
\bibitem [{\citenamefont {Cao}\ \emph {et~al.}(2018{\natexlab{a}})\citenamefont
  {Cao}, \citenamefont {Fatemi}, \citenamefont {Demir}, \citenamefont {Fang},
  \citenamefont {Tomarken}, \citenamefont {Luo}, \citenamefont
  {Sanchez-Yamagishi}, \citenamefont {Watanabe}, \citenamefont {Taniguchi},
  \citenamefont {Kaxiras}, \citenamefont {Ashoori},\ and\ \citenamefont
  {Jarillo-Herrero}}]{CaoMott2018}%
  \BibitemOpen
  \bibfield  {author} {\bibinfo {author} {\bibfnamefont {Y.}~\bibnamefont
  {Cao}}, \bibinfo {author} {\bibfnamefont {V.}~\bibnamefont {Fatemi}},
  \bibinfo {author} {\bibfnamefont {A.}~\bibnamefont {Demir}}, \bibinfo
  {author} {\bibfnamefont {S.}~\bibnamefont {Fang}}, \bibinfo {author}
  {\bibfnamefont {S.~L.}\ \bibnamefont {Tomarken}}, \bibinfo {author}
  {\bibfnamefont {J.~Y.}\ \bibnamefont {Luo}}, \bibinfo {author} {\bibfnamefont
  {J.~D.}\ \bibnamefont {Sanchez-Yamagishi}}, \bibinfo {author} {\bibfnamefont
  {K.}~\bibnamefont {Watanabe}}, \bibinfo {author} {\bibfnamefont
  {T.}~\bibnamefont {Taniguchi}}, \bibinfo {author} {\bibfnamefont
  {E.}~\bibnamefont {Kaxiras}}, \bibinfo {author} {\bibfnamefont {R.~C.}\
  \bibnamefont {Ashoori}}, \ and\ \bibinfo {author} {\bibfnamefont
  {P.}~\bibnamefont {Jarillo-Herrero}},\ }\href {\doibase 10.1038/nature26154}
  {\bibfield  {journal} {\bibinfo  {journal} {Nature}\ }\textbf {\bibinfo
  {volume} {556}},\ \bibinfo {pages} {80} (\bibinfo {year}
  {2018}{\natexlab{a}})}\BibitemShut {NoStop}%
\bibitem [{\citenamefont {Norman}(2016)}]{RevModPhys.88.041002}%
  \BibitemOpen
  \bibfield  {author} {\bibinfo {author} {\bibfnamefont {M.~R.}\ \bibnamefont
  {Norman}},\ }\href {\doibase 10.1103/RevModPhys.88.041002} {\bibfield
  {journal} {\bibinfo  {journal} {Rev. Mod. Phys.}\ }\textbf {\bibinfo {volume}
  {88}},\ \bibinfo {pages} {041002} (\bibinfo {year} {2016})}\BibitemShut
  {NoStop}%
\bibitem [{\citenamefont {Eich}\ \emph {et~al.}(2018)\citenamefont {Eich},
  \citenamefont {Pisoni}, \citenamefont {Pally}, \citenamefont {Overweg},
  \citenamefont {Kurzmann}, \citenamefont {Lee}, \citenamefont {Rickhaus},
  \citenamefont {Watanabe}, \citenamefont {Taniguchi}, \citenamefont
  {Ensslin},\ and\ \citenamefont {Ihn}}]{eich2018coupled}%
  \BibitemOpen
  \bibfield  {author} {\bibinfo {author} {\bibfnamefont {M.}~\bibnamefont
  {Eich}}, \bibinfo {author} {\bibfnamefont {R.}~\bibnamefont {Pisoni}},
  \bibinfo {author} {\bibfnamefont {A.}~\bibnamefont {Pally}}, \bibinfo
  {author} {\bibfnamefont {H.}~\bibnamefont {Overweg}}, \bibinfo {author}
  {\bibfnamefont {A.}~\bibnamefont {Kurzmann}}, \bibinfo {author}
  {\bibfnamefont {Y.}~\bibnamefont {Lee}}, \bibinfo {author} {\bibfnamefont
  {P.}~\bibnamefont {Rickhaus}}, \bibinfo {author} {\bibfnamefont
  {K.}~\bibnamefont {Watanabe}}, \bibinfo {author} {\bibfnamefont
  {T.}~\bibnamefont {Taniguchi}}, \bibinfo {author} {\bibfnamefont
  {K.}~\bibnamefont {Ensslin}}, \ and\ \bibinfo {author} {\bibfnamefont
  {T.}~\bibnamefont {Ihn}},\ }\href {\doibase 10.1021/acs.nanolett.8b01859}
  {\bibfield  {journal} {\bibinfo  {journal} {Nano Letters}\ }\textbf {\bibinfo
  {volume} {18}},\ \bibinfo {pages} {5042} (\bibinfo {year}
  {2018})}\BibitemShut {NoStop}%
\bibitem [{\citenamefont {Kelly}\ \emph {et~al.}(2016)\citenamefont {Kelly},
  \citenamefont {Gallagher},\ and\ \citenamefont
  {McQueen}}]{PhysRevX.6.041007}%
  \BibitemOpen
  \bibfield  {author} {\bibinfo {author} {\bibfnamefont {Z.~A.}\ \bibnamefont
  {Kelly}}, \bibinfo {author} {\bibfnamefont {M.~J.}\ \bibnamefont
  {Gallagher}}, \ and\ \bibinfo {author} {\bibfnamefont {T.~M.}\ \bibnamefont
  {McQueen}},\ }\href {\doibase 10.1103/PhysRevX.6.041007} {\bibfield
  {journal} {\bibinfo  {journal} {Phys. Rev. X}\ }\textbf {\bibinfo {volume}
  {6}},\ \bibinfo {pages} {041007} (\bibinfo {year} {2016})}\BibitemShut
  {NoStop}%
\bibitem [{\citenamefont {Banerjee}\ \emph {et~al.}(2016)\citenamefont
  {Banerjee}, \citenamefont {Bridges}, \citenamefont {Yan}, \citenamefont
  {Aczel}, \citenamefont {Li}, \citenamefont {Stone}, \citenamefont {Granroth},
  \citenamefont {Lumsden}, \citenamefont {Yiu}, \citenamefont {Knolle},
  \citenamefont {Bhattacharjee}, \citenamefont {Kovrizhin}, \citenamefont
  {Moessner}, \citenamefont {Tennant}, \citenamefont {Mandrus},\ and\
  \citenamefont {Nagler}}]{Banerjee2016}%
  \BibitemOpen
  \bibfield  {author} {\bibinfo {author} {\bibfnamefont {A.}~\bibnamefont
  {Banerjee}}, \bibinfo {author} {\bibfnamefont {C.~A.}\ \bibnamefont
  {Bridges}}, \bibinfo {author} {\bibfnamefont {J.-Q.}\ \bibnamefont {Yan}},
  \bibinfo {author} {\bibfnamefont {A.~A.}\ \bibnamefont {Aczel}}, \bibinfo
  {author} {\bibfnamefont {L.}~\bibnamefont {Li}}, \bibinfo {author}
  {\bibfnamefont {M.~B.}\ \bibnamefont {Stone}}, \bibinfo {author}
  {\bibfnamefont {G.~E.}\ \bibnamefont {Granroth}}, \bibinfo {author}
  {\bibfnamefont {M.~D.}\ \bibnamefont {Lumsden}}, \bibinfo {author}
  {\bibfnamefont {Y.}~\bibnamefont {Yiu}}, \bibinfo {author} {\bibfnamefont
  {J.}~\bibnamefont {Knolle}}, \bibinfo {author} {\bibfnamefont
  {S.}~\bibnamefont {Bhattacharjee}}, \bibinfo {author} {\bibfnamefont {D.~L.}\
  \bibnamefont {Kovrizhin}}, \bibinfo {author} {\bibfnamefont {R.}~\bibnamefont
  {Moessner}}, \bibinfo {author} {\bibfnamefont {D.~A.}\ \bibnamefont
  {Tennant}}, \bibinfo {author} {\bibfnamefont {D.~G.}\ \bibnamefont
  {Mandrus}}, \ and\ \bibinfo {author} {\bibfnamefont {S.~E.}\ \bibnamefont
  {Nagler}},\ }\href {\doibase 10.1038/nmat4604} {\bibfield  {journal}
  {\bibinfo  {journal} {Nature Materials}\ }\textbf {\bibinfo {volume} {15}},\
  \bibinfo {pages} {733} (\bibinfo {year} {2016})}\BibitemShut {NoStop}%
\bibitem [{\citenamefont {Sears}\ \emph {et~al.}(2015)\citenamefont {Sears},
  \citenamefont {Songvilay}, \citenamefont {Plumb}, \citenamefont {Clancy},
  \citenamefont {Qiu}, \citenamefont {Zhao}, \citenamefont {Parshall},\ and\
  \citenamefont {Kim}}]{PhysRevB.91.144420}%
  \BibitemOpen
  \bibfield  {author} {\bibinfo {author} {\bibfnamefont {J.~A.}\ \bibnamefont
  {Sears}}, \bibinfo {author} {\bibfnamefont {M.}~\bibnamefont {Songvilay}},
  \bibinfo {author} {\bibfnamefont {K.~W.}\ \bibnamefont {Plumb}}, \bibinfo
  {author} {\bibfnamefont {J.~P.}\ \bibnamefont {Clancy}}, \bibinfo {author}
  {\bibfnamefont {Y.}~\bibnamefont {Qiu}}, \bibinfo {author} {\bibfnamefont
  {Y.}~\bibnamefont {Zhao}}, \bibinfo {author} {\bibfnamefont {D.}~\bibnamefont
  {Parshall}}, \ and\ \bibinfo {author} {\bibfnamefont {Y.-J.}\ \bibnamefont
  {Kim}},\ }\href {\doibase 10.1103/PhysRevB.91.144420} {\bibfield  {journal}
  {\bibinfo  {journal} {Phys. Rev. B}\ }\textbf {\bibinfo {volume} {91}},\
  \bibinfo {pages} {144420} (\bibinfo {year} {2015})}\BibitemShut {NoStop}%
\bibitem [{\citenamefont {Guinea}\ \emph {et~al.}(2009)\citenamefont {Guinea},
  \citenamefont {Katsnelson},\ and\ \citenamefont {Geim}}]{Guinea2009}%
  \BibitemOpen
  \bibfield  {author} {\bibinfo {author} {\bibfnamefont {F.}~\bibnamefont
  {Guinea}}, \bibinfo {author} {\bibfnamefont {M.~I.}\ \bibnamefont
  {Katsnelson}}, \ and\ \bibinfo {author} {\bibfnamefont {A.~K.}\ \bibnamefont
  {Geim}},\ }\href {\doibase 10.1038/nphys1420} {\bibfield  {journal} {\bibinfo
   {journal} {Nature Physics}\ }\textbf {\bibinfo {volume} {6}},\ \bibinfo
  {pages} {30} (\bibinfo {year} {2009})}\BibitemShut {NoStop}%
\bibitem [{\citenamefont {Vozmediano}\ \emph {et~al.}(2010)\citenamefont
  {Vozmediano}, \citenamefont {Katsnelson},\ and\ \citenamefont
  {Guinea}}]{Voz2010}%
  \BibitemOpen
  \bibfield  {author} {\bibinfo {author} {\bibfnamefont {M.}~\bibnamefont
  {Vozmediano}}, \bibinfo {author} {\bibfnamefont {M.}~\bibnamefont
  {Katsnelson}}, \ and\ \bibinfo {author} {\bibfnamefont {F.}~\bibnamefont
  {Guinea}},\ }\href {\doibase 10.1016/j.physrep.2010.07.003} {\bibfield
  {journal} {\bibinfo  {journal} {Physics Reports}\ }\textbf {\bibinfo {volume}
  {496}},\ \bibinfo {pages} {109} (\bibinfo {year} {2010})}\BibitemShut
  {NoStop}%
\bibitem [{\citenamefont {Amorim}\ \emph {et~al.}(2016)\citenamefont {Amorim},
  \citenamefont {Cortijo}, \citenamefont {de~Juan}, \citenamefont {Grushin},
  \citenamefont {Guinea}, \citenamefont {Guti{\'{e}}rrez-Rubio}, \citenamefont
  {Ochoa}, \citenamefont {Parente}, \citenamefont {Rold{\'{a}}n}, \citenamefont
  {San-Jose}, \citenamefont {Schiefele}, \citenamefont {Sturla},\ and\
  \citenamefont {Vozmediano}}]{amorim2016novel}%
  \BibitemOpen
  \bibfield  {author} {\bibinfo {author} {\bibfnamefont {B.}~\bibnamefont
  {Amorim}}, \bibinfo {author} {\bibfnamefont {A.}~\bibnamefont {Cortijo}},
  \bibinfo {author} {\bibfnamefont {F.}~\bibnamefont {de~Juan}}, \bibinfo
  {author} {\bibfnamefont {A.}~\bibnamefont {Grushin}}, \bibinfo {author}
  {\bibfnamefont {F.}~\bibnamefont {Guinea}}, \bibinfo {author} {\bibfnamefont
  {A.}~\bibnamefont {Guti{\'{e}}rrez-Rubio}}, \bibinfo {author} {\bibfnamefont
  {H.}~\bibnamefont {Ochoa}}, \bibinfo {author} {\bibfnamefont
  {V.}~\bibnamefont {Parente}}, \bibinfo {author} {\bibfnamefont
  {R.}~\bibnamefont {Rold{\'{a}}n}}, \bibinfo {author} {\bibfnamefont
  {P.}~\bibnamefont {San-Jose}}, \bibinfo {author} {\bibfnamefont
  {J.}~\bibnamefont {Schiefele}}, \bibinfo {author} {\bibfnamefont
  {M.}~\bibnamefont {Sturla}}, \ and\ \bibinfo {author} {\bibfnamefont
  {M.}~\bibnamefont {Vozmediano}},\ }\href {\doibase
  10.1016/j.physrep.2015.12.006} {\bibfield  {journal} {\bibinfo  {journal}
  {Physics Reports}\ }\textbf {\bibinfo {volume} {617}},\ \bibinfo {pages} {1}
  (\bibinfo {year} {2016})}\BibitemShut {NoStop}%
\bibitem [{\citenamefont {Rozhkov}\ \emph {et~al.}(2016)\citenamefont
  {Rozhkov}, \citenamefont {Sboychakov}, \citenamefont {Rakhmanov},\ and\
  \citenamefont {Nori}}]{ROZHKOV20161}%
  \BibitemOpen
  \bibfield  {author} {\bibinfo {author} {\bibfnamefont {A.}~\bibnamefont
  {Rozhkov}}, \bibinfo {author} {\bibfnamefont {A.}~\bibnamefont {Sboychakov}},
  \bibinfo {author} {\bibfnamefont {A.}~\bibnamefont {Rakhmanov}}, \ and\
  \bibinfo {author} {\bibfnamefont {F.}~\bibnamefont {Nori}},\ }\href {\doibase
  https://doi.org/10.1016/j.physrep.2016.07.003} {\bibfield  {journal}
  {\bibinfo  {journal} {Physics Reports}\ }\textbf {\bibinfo {volume} {648}},\
  \bibinfo {pages} {1 } (\bibinfo {year} {2016})},\ \bibinfo {note} {electronic
  properties of graphene-based bilayer systems}\BibitemShut {NoStop}%
\bibitem [{\citenamefont {Sboychakov}\ \emph {et~al.}(2015)\citenamefont
  {Sboychakov}, \citenamefont {Rakhmanov}, \citenamefont {Rozhkov},\ and\
  \citenamefont {Nori}}]{PhysRevB.92.075402}%
  \BibitemOpen
  \bibfield  {author} {\bibinfo {author} {\bibfnamefont {A.~O.}\ \bibnamefont
  {Sboychakov}}, \bibinfo {author} {\bibfnamefont {A.~L.}\ \bibnamefont
  {Rakhmanov}}, \bibinfo {author} {\bibfnamefont {A.~V.}\ \bibnamefont
  {Rozhkov}}, \ and\ \bibinfo {author} {\bibfnamefont {F.}~\bibnamefont
  {Nori}},\ }\href {\doibase 10.1103/PhysRevB.92.075402} {\bibfield  {journal}
  {\bibinfo  {journal} {Phys. Rev. B}\ }\textbf {\bibinfo {volume} {92}},\
  \bibinfo {pages} {075402} (\bibinfo {year} {2015})}\BibitemShut {NoStop}%
\bibitem [{\citenamefont {L\"ofwander}\ \emph {et~al.}(2013)\citenamefont
  {L\"ofwander}, \citenamefont {San-Jose},\ and\ \citenamefont
  {Prada}}]{PhysRevB.87.205429}%
  \BibitemOpen
  \bibfield  {author} {\bibinfo {author} {\bibfnamefont {T.}~\bibnamefont
  {L\"ofwander}}, \bibinfo {author} {\bibfnamefont {P.}~\bibnamefont
  {San-Jose}}, \ and\ \bibinfo {author} {\bibfnamefont {E.}~\bibnamefont
  {Prada}},\ }\href {\doibase 10.1103/PhysRevB.87.205429} {\bibfield  {journal}
  {\bibinfo  {journal} {Phys. Rev. B}\ }\textbf {\bibinfo {volume} {87}},\
  \bibinfo {pages} {205429} (\bibinfo {year} {2013})}\BibitemShut {NoStop}%
\bibitem [{\citenamefont {Wright}\ and\ \citenamefont
  {Hyart}(2011)}]{wright2011robust}%
  \BibitemOpen
  \bibfield  {author} {\bibinfo {author} {\bibfnamefont {A.~R.}\ \bibnamefont
  {Wright}}\ and\ \bibinfo {author} {\bibfnamefont {T.}~\bibnamefont {Hyart}},\
  }\href {\doibase 10.1063/1.3601851} {\bibfield  {journal} {\bibinfo
  {journal} {Applied Physics Letters}\ }\textbf {\bibinfo {volume} {98}},\
  \bibinfo {pages} {251902} (\bibinfo {year} {2011})},\ \Eprint
  {http://arxiv.org/abs/https://doi.org/10.1063/1.3601851}
  {https://doi.org/10.1063/1.3601851} \BibitemShut {NoStop}%
\bibitem [{\citenamefont {Vaezi}\ \emph {et~al.}(2013)\citenamefont {Vaezi},
  \citenamefont {Liang}, \citenamefont {Ngai}, \citenamefont {Yang},\ and\
  \citenamefont {Kim}}]{PhysRevX.3.021018}%
  \BibitemOpen
  \bibfield  {author} {\bibinfo {author} {\bibfnamefont {A.}~\bibnamefont
  {Vaezi}}, \bibinfo {author} {\bibfnamefont {Y.}~\bibnamefont {Liang}},
  \bibinfo {author} {\bibfnamefont {D.~H.}\ \bibnamefont {Ngai}}, \bibinfo
  {author} {\bibfnamefont {L.}~\bibnamefont {Yang}}, \ and\ \bibinfo {author}
  {\bibfnamefont {E.-A.}\ \bibnamefont {Kim}},\ }\href {\doibase
  10.1103/PhysRevX.3.021018} {\bibfield  {journal} {\bibinfo  {journal} {Phys.
  Rev. X}\ }\textbf {\bibinfo {volume} {3}},\ \bibinfo {pages} {021018}
  (\bibinfo {year} {2013})}\BibitemShut {NoStop}%
\bibitem [{\citenamefont {Jung}\ \emph {et~al.}(2014)\citenamefont {Jung},
  \citenamefont {Raoux}, \citenamefont {Qiao},\ and\ \citenamefont
  {MacDonald}}]{PhysRevB.89.205414}%
  \BibitemOpen
  \bibfield  {author} {\bibinfo {author} {\bibfnamefont {J.}~\bibnamefont
  {Jung}}, \bibinfo {author} {\bibfnamefont {A.}~\bibnamefont {Raoux}},
  \bibinfo {author} {\bibfnamefont {Z.}~\bibnamefont {Qiao}}, \ and\ \bibinfo
  {author} {\bibfnamefont {A.~H.}\ \bibnamefont {MacDonald}},\ }\href {\doibase
  10.1103/PhysRevB.89.205414} {\bibfield  {journal} {\bibinfo  {journal} {Phys.
  Rev. B}\ }\textbf {\bibinfo {volume} {89}},\ \bibinfo {pages} {205414}
  (\bibinfo {year} {2014})}\BibitemShut {NoStop}%
\bibitem [{\citenamefont {van Wijk}\ \emph {et~al.}(2015)\citenamefont {van
  Wijk}, \citenamefont {Schuring}, \citenamefont {Katsnelson},\ and\
  \citenamefont {Fasolino}}]{van2015relaxation}%
  \BibitemOpen
  \bibfield  {author} {\bibinfo {author} {\bibfnamefont {M.~M.}\ \bibnamefont
  {van Wijk}}, \bibinfo {author} {\bibfnamefont {A.}~\bibnamefont {Schuring}},
  \bibinfo {author} {\bibfnamefont {M.~I.}\ \bibnamefont {Katsnelson}}, \ and\
  \bibinfo {author} {\bibfnamefont {A.}~\bibnamefont {Fasolino}},\ }\href
  {\doibase 10.1088/2053-1583/2/3/034010} {\bibfield  {journal} {\bibinfo
  {journal} {2D Materials}\ }\textbf {\bibinfo {volume} {2}},\ \bibinfo {pages}
  {034010} (\bibinfo {year} {2015})}\BibitemShut {NoStop}%
\bibitem [{\citenamefont {Cherkez}\ \emph {et~al.}(2015)\citenamefont
  {Cherkez}, \citenamefont {de~Laissardi\`ere}, \citenamefont {Mallet},\ and\
  \citenamefont {Veuillen}}]{PhysRevB.91.155428}%
  \BibitemOpen
  \bibfield  {author} {\bibinfo {author} {\bibfnamefont {V.}~\bibnamefont
  {Cherkez}}, \bibinfo {author} {\bibfnamefont {G.~T.}\ \bibnamefont
  {de~Laissardi\`ere}}, \bibinfo {author} {\bibfnamefont {P.}~\bibnamefont
  {Mallet}}, \ and\ \bibinfo {author} {\bibfnamefont {J.-Y.}\ \bibnamefont
  {Veuillen}},\ }\href {\doibase 10.1103/PhysRevB.91.155428} {\bibfield
  {journal} {\bibinfo  {journal} {Phys. Rev. B}\ }\textbf {\bibinfo {volume}
  {91}},\ \bibinfo {pages} {155428} (\bibinfo {year} {2015})}\BibitemShut
  {NoStop}%
\bibitem [{\citenamefont {Gargiulo}\ and\ \citenamefont
  {Yazyev}(2017)}]{gargiulo2017structural}%
  \BibitemOpen
  \bibfield  {author} {\bibinfo {author} {\bibfnamefont {F.}~\bibnamefont
  {Gargiulo}}\ and\ \bibinfo {author} {\bibfnamefont {O.~V.}\ \bibnamefont
  {Yazyev}},\ }\href {\doibase 10.1088/2053-1583/aa9640} {\bibfield  {journal}
  {\bibinfo  {journal} {2D Materials}\ }\textbf {\bibinfo {volume} {5}},\
  \bibinfo {pages} {015019} (\bibinfo {year} {2017})}\BibitemShut {NoStop}%
\bibitem [{\citenamefont {Fleischmann}\ \emph {et~al.}(2018)\citenamefont
  {Fleischmann}, \citenamefont {Gupta}, \citenamefont {Weckbecker},
  \citenamefont {Landgraf}, \citenamefont {Pankratov}, \citenamefont {Meded},\
  and\ \citenamefont {Shallcross}}]{fleischmann2018moir}%
  \BibitemOpen
  \bibfield  {author} {\bibinfo {author} {\bibfnamefont {M.}~\bibnamefont
  {Fleischmann}}, \bibinfo {author} {\bibfnamefont {R.}~\bibnamefont {Gupta}},
  \bibinfo {author} {\bibfnamefont {D.}~\bibnamefont {Weckbecker}}, \bibinfo
  {author} {\bibfnamefont {W.}~\bibnamefont {Landgraf}}, \bibinfo {author}
  {\bibfnamefont {O.}~\bibnamefont {Pankratov}}, \bibinfo {author}
  {\bibfnamefont {V.}~\bibnamefont {Meded}}, \ and\ \bibinfo {author}
  {\bibfnamefont {S.}~\bibnamefont {Shallcross}},\ }\href {\doibase
  10.1103/PhysRevB.97.205128} {\bibfield  {journal} {\bibinfo  {journal} {Phys.
  Rev. B}\ }\textbf {\bibinfo {volume} {97}},\ \bibinfo {pages} {205128}
  (\bibinfo {year} {2018})}\BibitemShut {NoStop}%
\bibitem [{\citenamefont {Andelkovic}\ \emph {et~al.}(2018)\citenamefont
  {Andelkovic}, \citenamefont {Covaci},\ and\ \citenamefont
  {Peeters}}]{PhysRevMaterials.2.034004}%
  \BibitemOpen
  \bibfield  {author} {\bibinfo {author} {\bibfnamefont {M.}~\bibnamefont
  {Andelkovic}}, \bibinfo {author} {\bibfnamefont {L.}~\bibnamefont {Covaci}},
  \ and\ \bibinfo {author} {\bibfnamefont {F.~M.}\ \bibnamefont {Peeters}},\
  }\href {\doibase 10.1103/PhysRevMaterials.2.034004} {\bibfield  {journal}
  {\bibinfo  {journal} {Phys. Rev. Materials}\ }\textbf {\bibinfo {volume}
  {2}},\ \bibinfo {pages} {034004} (\bibinfo {year} {2018})}\BibitemShut
  {NoStop}%
\bibitem [{\citenamefont {Lee}\ \emph {et~al.}(2011)\citenamefont {Lee},
  \citenamefont {Riedl}, \citenamefont {Beringer}, \citenamefont {Castro~Neto},
  \citenamefont {von Klitzing}, \citenamefont {Starke},\ and\ \citenamefont
  {Smet}}]{PhysRevLett.107.216602}%
  \BibitemOpen
  \bibfield  {author} {\bibinfo {author} {\bibfnamefont {D.~S.}\ \bibnamefont
  {Lee}}, \bibinfo {author} {\bibfnamefont {C.}~\bibnamefont {Riedl}}, \bibinfo
  {author} {\bibfnamefont {T.}~\bibnamefont {Beringer}}, \bibinfo {author}
  {\bibfnamefont {A.~H.}\ \bibnamefont {Castro~Neto}}, \bibinfo {author}
  {\bibfnamefont {K.}~\bibnamefont {von Klitzing}}, \bibinfo {author}
  {\bibfnamefont {U.}~\bibnamefont {Starke}}, \ and\ \bibinfo {author}
  {\bibfnamefont {J.~H.}\ \bibnamefont {Smet}},\ }\href {\doibase
  10.1103/PhysRevLett.107.216602} {\bibfield  {journal} {\bibinfo  {journal}
  {Phys. Rev. Lett.}\ }\textbf {\bibinfo {volume} {107}},\ \bibinfo {pages}
  {216602} (\bibinfo {year} {2011})}\BibitemShut {NoStop}%
\bibitem [{\citenamefont {San-Jose}\ \emph {et~al.}(2012)\citenamefont
  {San-Jose}, \citenamefont {Gonz\'alez},\ and\ \citenamefont
  {Guinea}}]{PhysRevLett.108.216802}%
  \BibitemOpen
  \bibfield  {author} {\bibinfo {author} {\bibfnamefont {P.}~\bibnamefont
  {San-Jose}}, \bibinfo {author} {\bibfnamefont {J.}~\bibnamefont
  {Gonz\'alez}}, \ and\ \bibinfo {author} {\bibfnamefont {F.}~\bibnamefont
  {Guinea}},\ }\href {\doibase 10.1103/PhysRevLett.108.216802} {\bibfield
  {journal} {\bibinfo  {journal} {Phys. Rev. Lett.}\ }\textbf {\bibinfo
  {volume} {108}},\ \bibinfo {pages} {216802} (\bibinfo {year}
  {2012})}\BibitemShut {NoStop}%
\bibitem [{\citenamefont {Gonzalez-Arraga}\ \emph {et~al.}(2017)\citenamefont
  {Gonzalez-Arraga}, \citenamefont {Lado}, \citenamefont {Guinea},\ and\
  \citenamefont {San-Jose}}]{PhysRevLett.119.107201}%
  \BibitemOpen
  \bibfield  {author} {\bibinfo {author} {\bibfnamefont {L.~A.}\ \bibnamefont
  {Gonzalez-Arraga}}, \bibinfo {author} {\bibfnamefont {J.~L.}\ \bibnamefont
  {Lado}}, \bibinfo {author} {\bibfnamefont {F.}~\bibnamefont {Guinea}}, \ and\
  \bibinfo {author} {\bibfnamefont {P.}~\bibnamefont {San-Jose}},\ }\href
  {\doibase 10.1103/PhysRevLett.119.107201} {\bibfield  {journal} {\bibinfo
  {journal} {Phys. Rev. Lett.}\ }\textbf {\bibinfo {volume} {119}},\ \bibinfo
  {pages} {107201} (\bibinfo {year} {2017})}\BibitemShut {NoStop}%
\bibitem [{\citenamefont {Esquinazi}\ \emph {et~al.}(2014)\citenamefont
  {Esquinazi}, \citenamefont {Heikkil\"{a}}, \citenamefont {Lysogorskiy},
  \citenamefont {Tayurskii},\ and\ \citenamefont
  {Volovik}}]{esquinazi2014superconductivity}%
  \BibitemOpen
  \bibfield  {author} {\bibinfo {author} {\bibfnamefont {P.}~\bibnamefont
  {Esquinazi}}, \bibinfo {author} {\bibfnamefont {T.~T.}\ \bibnamefont
  {Heikkil\"{a}}}, \bibinfo {author} {\bibfnamefont {Y.~V.}\ \bibnamefont
  {Lysogorskiy}}, \bibinfo {author} {\bibfnamefont {D.~A.}\ \bibnamefont
  {Tayurskii}}, \ and\ \bibinfo {author} {\bibfnamefont {G.~E.}\ \bibnamefont
  {Volovik}},\ }\href {\doibase 10.1134/s0021364014170056} {\bibfield
  {journal} {\bibinfo  {journal} {{JETP} Letters}\ }\textbf {\bibinfo {volume}
  {100}},\ \bibinfo {pages} {336} (\bibinfo {year} {2014})}\BibitemShut
  {NoStop}%
\bibitem [{\citenamefont {Huang}\ \emph {et~al.}(2017)\citenamefont {Huang},
  \citenamefont {Yankowitz}, \citenamefont {Chattrakun}, \citenamefont
  {Sandhu},\ and\ \citenamefont {LeRoy}}]{huang2017evolution}%
  \BibitemOpen
  \bibfield  {author} {\bibinfo {author} {\bibfnamefont {S.}~\bibnamefont
  {Huang}}, \bibinfo {author} {\bibfnamefont {M.}~\bibnamefont {Yankowitz}},
  \bibinfo {author} {\bibfnamefont {K.}~\bibnamefont {Chattrakun}}, \bibinfo
  {author} {\bibfnamefont {A.}~\bibnamefont {Sandhu}}, \ and\ \bibinfo {author}
  {\bibfnamefont {B.~J.}\ \bibnamefont {LeRoy}},\ }\href {\doibase
  10.1038/s41598-017-07580-3} {\bibfield  {journal} {\bibinfo  {journal}
  {Scientific Reports}\ }\textbf {\bibinfo {volume} {7}} (\bibinfo {year}
  {2017}),\ 10.1038/s41598-017-07580-3}\BibitemShut {NoStop}%
\bibitem [{\citenamefont {Bistritzer}\ and\ \citenamefont
  {MacDonald}(2011)}]{bistritzer2011moire}%
  \BibitemOpen
  \bibfield  {author} {\bibinfo {author} {\bibfnamefont {R.}~\bibnamefont
  {Bistritzer}}\ and\ \bibinfo {author} {\bibfnamefont {A.~H.}\ \bibnamefont
  {MacDonald}},\ }\href {\doibase 10.1073/pnas.1108174108} {\bibfield
  {journal} {\bibinfo  {journal} {Proceedings of the National Academy of
  Sciences}\ }\textbf {\bibinfo {volume} {108}},\ \bibinfo {pages} {12233}
  (\bibinfo {year} {2011})}\BibitemShut {NoStop}%
\bibitem [{\citenamefont {Cao}\ \emph {et~al.}(2016)\citenamefont {Cao},
  \citenamefont {Luo}, \citenamefont {Fatemi}, \citenamefont {Fang},
  \citenamefont {Sanchez-Yamagishi}, \citenamefont {Watanabe}, \citenamefont
  {Taniguchi}, \citenamefont {Kaxiras},\ and\ \citenamefont
  {Jarillo-Herrero}}]{PhysRevLett.117.116804}%
  \BibitemOpen
  \bibfield  {author} {\bibinfo {author} {\bibfnamefont {Y.}~\bibnamefont
  {Cao}}, \bibinfo {author} {\bibfnamefont {J.~Y.}\ \bibnamefont {Luo}},
  \bibinfo {author} {\bibfnamefont {V.}~\bibnamefont {Fatemi}}, \bibinfo
  {author} {\bibfnamefont {S.}~\bibnamefont {Fang}}, \bibinfo {author}
  {\bibfnamefont {J.~D.}\ \bibnamefont {Sanchez-Yamagishi}}, \bibinfo {author}
  {\bibfnamefont {K.}~\bibnamefont {Watanabe}}, \bibinfo {author}
  {\bibfnamefont {T.}~\bibnamefont {Taniguchi}}, \bibinfo {author}
  {\bibfnamefont {E.}~\bibnamefont {Kaxiras}}, \ and\ \bibinfo {author}
  {\bibfnamefont {P.}~\bibnamefont {Jarillo-Herrero}},\ }\href {\doibase
  10.1103/PhysRevLett.117.116804} {\bibfield  {journal} {\bibinfo  {journal}
  {Phys. Rev. Lett.}\ }\textbf {\bibinfo {volume} {117}},\ \bibinfo {pages}
  {116804} (\bibinfo {year} {2016})}\BibitemShut {NoStop}%
\bibitem [{\citenamefont {Lopes~dos Santos}\ \emph {et~al.}(2012)\citenamefont
  {Lopes~dos Santos}, \citenamefont {Peres},\ and\ \citenamefont
  {Castro~Neto}}]{PhysRevB.86.155449}%
  \BibitemOpen
  \bibfield  {author} {\bibinfo {author} {\bibfnamefont {J.~M.~B.}\
  \bibnamefont {Lopes~dos Santos}}, \bibinfo {author} {\bibfnamefont
  {N.~M.~R.}\ \bibnamefont {Peres}}, \ and\ \bibinfo {author} {\bibfnamefont
  {A.~H.}\ \bibnamefont {Castro~Neto}},\ }\href {\doibase
  10.1103/PhysRevB.86.155449} {\bibfield  {journal} {\bibinfo  {journal} {Phys.
  Rev. B}\ }\textbf {\bibinfo {volume} {86}},\ \bibinfo {pages} {155449}
  (\bibinfo {year} {2012})}\BibitemShut {NoStop}%
\bibitem [{\citenamefont {Finocchiaro}\ \emph {et~al.}(2017)\citenamefont
  {Finocchiaro}, \citenamefont {Guinea},\ and\ \citenamefont
  {San-Jose}}]{finocchiaro2017quantum}%
  \BibitemOpen
  \bibfield  {author} {\bibinfo {author} {\bibfnamefont {F.}~\bibnamefont
  {Finocchiaro}}, \bibinfo {author} {\bibfnamefont {F.}~\bibnamefont {Guinea}},
  \ and\ \bibinfo {author} {\bibfnamefont {P.}~\bibnamefont {San-Jose}},\
  }\href {\doibase 10.1088/2053-1583/aa5265} {\bibfield  {journal} {\bibinfo
  {journal} {2D Materials}\ }\textbf {\bibinfo {volume} {4}},\ \bibinfo {pages}
  {025027} (\bibinfo {year} {2017})}\BibitemShut {NoStop}%
\bibitem [{\citenamefont {San-Jose}\ and\ \citenamefont
  {Prada}(2013)}]{PhysRevB.88.121408}%
  \BibitemOpen
  \bibfield  {author} {\bibinfo {author} {\bibfnamefont {P.}~\bibnamefont
  {San-Jose}}\ and\ \bibinfo {author} {\bibfnamefont {E.}~\bibnamefont
  {Prada}},\ }\href {\doibase 10.1103/PhysRevB.88.121408} {\bibfield  {journal}
  {\bibinfo  {journal} {Phys. Rev. B}\ }\textbf {\bibinfo {volume} {88}},\
  \bibinfo {pages} {121408} (\bibinfo {year} {2013})}\BibitemShut {NoStop}%
\bibitem [{\citenamefont {{Rickhaus}}\ \emph {et~al.}(2018)\citenamefont
  {{Rickhaus}}, \citenamefont {{Wallbank}}, \citenamefont {{Slizovskiy}},
  \citenamefont {{Pisoni}}, \citenamefont {{Overweg}}, \citenamefont {{Lee}},
  \citenamefont {{Eich}}, \citenamefont {{Liu}}, \citenamefont {{Watanabe}},
  \citenamefont {{Taniguchi}}, \citenamefont {{Fal'ko}}, \citenamefont
  {{Ihn}},\ and\ \citenamefont {{Ensslin}}}]{rickhaus2018transport}%
  \BibitemOpen
  \bibfield  {author} {\bibinfo {author} {\bibfnamefont {P.}~\bibnamefont
  {{Rickhaus}}}, \bibinfo {author} {\bibfnamefont {J.}~\bibnamefont
  {{Wallbank}}}, \bibinfo {author} {\bibfnamefont {S.}~\bibnamefont
  {{Slizovskiy}}}, \bibinfo {author} {\bibfnamefont {R.}~\bibnamefont
  {{Pisoni}}}, \bibinfo {author} {\bibfnamefont {H.}~\bibnamefont {{Overweg}}},
  \bibinfo {author} {\bibfnamefont {Y.}~\bibnamefont {{Lee}}}, \bibinfo
  {author} {\bibfnamefont {M.}~\bibnamefont {{Eich}}}, \bibinfo {author}
  {\bibfnamefont {M.-H.}\ \bibnamefont {{Liu}}}, \bibinfo {author}
  {\bibfnamefont {K.}~\bibnamefont {{Watanabe}}}, \bibinfo {author}
  {\bibfnamefont {T.}~\bibnamefont {{Taniguchi}}}, \bibinfo {author}
  {\bibfnamefont {V.}~\bibnamefont {{Fal'ko}}}, \bibinfo {author}
  {\bibfnamefont {T.}~\bibnamefont {{Ihn}}}, \ and\ \bibinfo {author}
  {\bibfnamefont {K.}~\bibnamefont {{Ensslin}}},\ }\href@noop {} {\bibfield
  {journal} {\bibinfo  {journal} {ArXiv e-prints}\ } (\bibinfo {year}
  {2018})},\ \Eprint {http://arxiv.org/abs/1802.07317} {arXiv:1802.07317
  [cond-mat.mes-hall]} \BibitemShut {NoStop}%
\bibitem [{\citenamefont {Huang}\ \emph {et~al.}(2018)\citenamefont {Huang},
  \citenamefont {Kim}, \citenamefont {Efimkin}, \citenamefont {Lovorn},
  \citenamefont {Taniguchi}, \citenamefont {Watanabe}, \citenamefont
  {MacDonald}, \citenamefont {Tutuc},\ and\ \citenamefont
  {LeRoy}}]{huang2018emergence}%
  \BibitemOpen
  \bibfield  {author} {\bibinfo {author} {\bibfnamefont {S.}~\bibnamefont
  {Huang}}, \bibinfo {author} {\bibfnamefont {K.}~\bibnamefont {Kim}}, \bibinfo
  {author} {\bibfnamefont {D.~K.}\ \bibnamefont {Efimkin}}, \bibinfo {author}
  {\bibfnamefont {T.}~\bibnamefont {Lovorn}}, \bibinfo {author} {\bibfnamefont
  {T.}~\bibnamefont {Taniguchi}}, \bibinfo {author} {\bibfnamefont
  {K.}~\bibnamefont {Watanabe}}, \bibinfo {author} {\bibfnamefont {A.~H.}\
  \bibnamefont {MacDonald}}, \bibinfo {author} {\bibfnamefont {E.}~\bibnamefont
  {Tutuc}}, \ and\ \bibinfo {author} {\bibfnamefont {B.~J.}\ \bibnamefont
  {LeRoy}},\ }\href {\doibase 10.1103/PhysRevLett.121.037702} {\bibfield
  {journal} {\bibinfo  {journal} {Phys. Rev. Lett.}\ }\textbf {\bibinfo
  {volume} {121}},\ \bibinfo {pages} {037702} (\bibinfo {year}
  {2018})}\BibitemShut {NoStop}%
\bibitem [{\citenamefont {Su\'arez~Morell}\ \emph {et~al.}(2010)\citenamefont
  {Su\'arez~Morell}, \citenamefont {Correa}, \citenamefont {Vargas},
  \citenamefont {Pacheco},\ and\ \citenamefont
  {Barticevic}}]{PhysRevB.82.121407}%
  \BibitemOpen
  \bibfield  {author} {\bibinfo {author} {\bibfnamefont {E.}~\bibnamefont
  {Su\'arez~Morell}}, \bibinfo {author} {\bibfnamefont {J.~D.}\ \bibnamefont
  {Correa}}, \bibinfo {author} {\bibfnamefont {P.}~\bibnamefont {Vargas}},
  \bibinfo {author} {\bibfnamefont {M.}~\bibnamefont {Pacheco}}, \ and\
  \bibinfo {author} {\bibfnamefont {Z.}~\bibnamefont {Barticevic}},\ }\href
  {\doibase 10.1103/PhysRevB.82.121407} {\bibfield  {journal} {\bibinfo
  {journal} {Phys. Rev. B}\ }\textbf {\bibinfo {volume} {82}},\ \bibinfo
  {pages} {121407} (\bibinfo {year} {2010})}\BibitemShut {NoStop}%
\bibitem [{\citenamefont {Trambly~de Laissardi\`ere}\ \emph
  {et~al.}(2012)\citenamefont {Trambly~de Laissardi\`ere}, \citenamefont
  {Mayou},\ and\ \citenamefont {Magaud}}]{PhysRevB.86.125413}%
  \BibitemOpen
  \bibfield  {author} {\bibinfo {author} {\bibfnamefont {G.}~\bibnamefont
  {Trambly~de Laissardi\`ere}}, \bibinfo {author} {\bibfnamefont
  {D.}~\bibnamefont {Mayou}}, \ and\ \bibinfo {author} {\bibfnamefont
  {L.}~\bibnamefont {Magaud}},\ }\href {\doibase 10.1103/PhysRevB.86.125413}
  {\bibfield  {journal} {\bibinfo  {journal} {Phys. Rev. B}\ }\textbf {\bibinfo
  {volume} {86}},\ \bibinfo {pages} {125413} (\bibinfo {year}
  {2012})}\BibitemShut {NoStop}%
\bibitem [{\citenamefont {Kim}\ \emph {et~al.}(2017)\citenamefont {Kim},
  \citenamefont {DaSilva}, \citenamefont {Huang}, \citenamefont {Fallahazad},
  \citenamefont {Larentis}, \citenamefont {Taniguchi}, \citenamefont
  {Watanabe}, \citenamefont {LeRoy}, \citenamefont {MacDonald},\ and\
  \citenamefont {Tutuc}}]{kim2017tunable}%
  \BibitemOpen
  \bibfield  {author} {\bibinfo {author} {\bibfnamefont {K.}~\bibnamefont
  {Kim}}, \bibinfo {author} {\bibfnamefont {A.}~\bibnamefont {DaSilva}},
  \bibinfo {author} {\bibfnamefont {S.}~\bibnamefont {Huang}}, \bibinfo
  {author} {\bibfnamefont {B.}~\bibnamefont {Fallahazad}}, \bibinfo {author}
  {\bibfnamefont {S.}~\bibnamefont {Larentis}}, \bibinfo {author}
  {\bibfnamefont {T.}~\bibnamefont {Taniguchi}}, \bibinfo {author}
  {\bibfnamefont {K.}~\bibnamefont {Watanabe}}, \bibinfo {author}
  {\bibfnamefont {B.~J.}\ \bibnamefont {LeRoy}}, \bibinfo {author}
  {\bibfnamefont {A.~H.}\ \bibnamefont {MacDonald}}, \ and\ \bibinfo {author}
  {\bibfnamefont {E.}~\bibnamefont {Tutuc}},\ }\href {\doibase
  10.1073/pnas.1620140114} {\bibfield  {journal} {\bibinfo  {journal}
  {Proceedings of the National Academy of Sciences}\ }\textbf {\bibinfo
  {volume} {114}},\ \bibinfo {pages} {3364} (\bibinfo {year}
  {2017})}\BibitemShut {NoStop}%
\bibitem [{\citenamefont {Cao}\ \emph {et~al.}(2018{\natexlab{b}})\citenamefont
  {Cao}, \citenamefont {Fatemi}, \citenamefont {Fang}, \citenamefont
  {Watanabe}, \citenamefont {Taniguchi}, \citenamefont {Kaxiras},\ and\
  \citenamefont {Jarillo-Herrero}}]{cao2018unconventional}%
  \BibitemOpen
  \bibfield  {author} {\bibinfo {author} {\bibfnamefont {Y.}~\bibnamefont
  {Cao}}, \bibinfo {author} {\bibfnamefont {V.}~\bibnamefont {Fatemi}},
  \bibinfo {author} {\bibfnamefont {S.}~\bibnamefont {Fang}}, \bibinfo {author}
  {\bibfnamefont {K.}~\bibnamefont {Watanabe}}, \bibinfo {author}
  {\bibfnamefont {T.}~\bibnamefont {Taniguchi}}, \bibinfo {author}
  {\bibfnamefont {E.}~\bibnamefont {Kaxiras}}, \ and\ \bibinfo {author}
  {\bibfnamefont {P.}~\bibnamefont {Jarillo-Herrero}},\ }\href {\doibase
  10.1038/nature26160} {\bibfield  {journal} {\bibinfo  {journal} {Nature}\
  }\textbf {\bibinfo {volume} {556}},\ \bibinfo {pages} {43} (\bibinfo {year}
  {2018}{\natexlab{b}})}\BibitemShut {NoStop}%
\bibitem [{\citenamefont {McCann}\ and\ \citenamefont
  {Koshino}(2013)}]{mccann2013electronic}%
  \BibitemOpen
  \bibfield  {author} {\bibinfo {author} {\bibfnamefont {E.}~\bibnamefont
  {McCann}}\ and\ \bibinfo {author} {\bibfnamefont {M.}~\bibnamefont
  {Koshino}},\ }\href {\doibase 10.1088/0034-4885/76/5/056503} {\bibfield
  {journal} {\bibinfo  {journal} {Reports on Progress in Physics}\ }\textbf
  {\bibinfo {volume} {76}},\ \bibinfo {pages} {056503} (\bibinfo {year}
  {2013})}\BibitemShut {NoStop}%
\bibitem [{\citenamefont {Sun}\ \emph {et~al.}(2010{\natexlab{a}})\citenamefont
  {Sun}, \citenamefont {Fertig},\ and\ \citenamefont
  {Brey}}]{PhysRevLett.105.156801}%
  \BibitemOpen
  \bibfield  {author} {\bibinfo {author} {\bibfnamefont {J.}~\bibnamefont
  {Sun}}, \bibinfo {author} {\bibfnamefont {H.~A.}\ \bibnamefont {Fertig}}, \
  and\ \bibinfo {author} {\bibfnamefont {L.}~\bibnamefont {Brey}},\ }\href
  {\doibase 10.1103/PhysRevLett.105.156801} {\bibfield  {journal} {\bibinfo
  {journal} {Phys. Rev. Lett.}\ }\textbf {\bibinfo {volume} {105}},\ \bibinfo
  {pages} {156801} (\bibinfo {year} {2010}{\natexlab{a}})}\BibitemShut
  {NoStop}%
\bibitem [{\citenamefont {Ando}(2005{\natexlab{a}})}]{ando2005theory}%
  \BibitemOpen
  \bibfield  {author} {\bibinfo {author} {\bibfnamefont {T.}~\bibnamefont
  {Ando}},\ }\href {\doibase 10.1143/jpsj.74.777} {\bibfield  {journal}
  {\bibinfo  {journal} {Journal of the Physical Society of Japan}\ }\textbf
  {\bibinfo {volume} {74}},\ \bibinfo {pages} {777} (\bibinfo {year}
  {2005}{\natexlab{a}})}\BibitemShut {NoStop}%
\bibitem [{\citenamefont {Yankowitz}\ \emph {et~al.}(2012)\citenamefont
  {Yankowitz}, \citenamefont {Xue}, \citenamefont {Cormode}, \citenamefont
  {Sanchez-Yamagishi}, \citenamefont {Watanabe}, \citenamefont {Taniguchi},
  \citenamefont {Jarillo-Herrero}, \citenamefont {Jacquod},\ and\ \citenamefont
  {LeRoy}}]{Yankowitz2012}%
  \BibitemOpen
  \bibfield  {author} {\bibinfo {author} {\bibfnamefont {M.}~\bibnamefont
  {Yankowitz}}, \bibinfo {author} {\bibfnamefont {J.}~\bibnamefont {Xue}},
  \bibinfo {author} {\bibfnamefont {D.}~\bibnamefont {Cormode}}, \bibinfo
  {author} {\bibfnamefont {J.~D.}\ \bibnamefont {Sanchez-Yamagishi}}, \bibinfo
  {author} {\bibfnamefont {K.}~\bibnamefont {Watanabe}}, \bibinfo {author}
  {\bibfnamefont {T.}~\bibnamefont {Taniguchi}}, \bibinfo {author}
  {\bibfnamefont {P.}~\bibnamefont {Jarillo-Herrero}}, \bibinfo {author}
  {\bibfnamefont {P.}~\bibnamefont {Jacquod}}, \ and\ \bibinfo {author}
  {\bibfnamefont {B.~J.}\ \bibnamefont {LeRoy}},\ }\href
  {http://dx.doi.org/10.1038/nphys2272 http://10.0.4.14/nphys2272
  https://www.nature.com/articles/nphys2272{\#}supplementary-information}
  {\bibfield  {journal} {\bibinfo  {journal} {Nature Physics}\ }\textbf
  {\bibinfo {volume} {8}},\ \bibinfo {pages} {382} (\bibinfo {year}
  {2012})}\BibitemShut {NoStop}%
\bibitem [{\citenamefont {{Chen}}\ \emph {et~al.}(2018)\citenamefont {{Chen}},
  \citenamefont {{Jiang}}, \citenamefont {{Wu}}, \citenamefont {{Lv}},
  \citenamefont {{Li}}, \citenamefont {{Watanabe}}, \citenamefont
  {{Taniguchi}}, \citenamefont {{Shi}}, \citenamefont {{Zhang}},\ and\
  \citenamefont {{Wang}}}]{Chen2018}%
  \BibitemOpen
  \bibfield  {author} {\bibinfo {author} {\bibfnamefont {G.}~\bibnamefont
  {{Chen}}}, \bibinfo {author} {\bibfnamefont {L.}~\bibnamefont {{Jiang}}},
  \bibinfo {author} {\bibfnamefont {S.}~\bibnamefont {{Wu}}}, \bibinfo {author}
  {\bibfnamefont {B.}~\bibnamefont {{Lv}}}, \bibinfo {author} {\bibfnamefont
  {H.}~\bibnamefont {{Li}}}, \bibinfo {author} {\bibfnamefont {K.}~\bibnamefont
  {{Watanabe}}}, \bibinfo {author} {\bibfnamefont {T.}~\bibnamefont
  {{Taniguchi}}}, \bibinfo {author} {\bibfnamefont {Z.}~\bibnamefont {{Shi}}},
  \bibinfo {author} {\bibfnamefont {Y.}~\bibnamefont {{Zhang}}}, \ and\
  \bibinfo {author} {\bibfnamefont {F.}~\bibnamefont {{Wang}}},\ }\href@noop {}
  {\bibfield  {journal} {\bibinfo  {journal} {ArXiv e-prints}\ } (\bibinfo
  {year} {2018})},\ \Eprint {http://arxiv.org/abs/1803.01985} {arXiv:1803.01985
  [cond-mat.mes-hall]} \BibitemShut {NoStop}%
\bibitem [{\citenamefont {Brihuega}\ \emph {et~al.}(2012)\citenamefont
  {Brihuega}, \citenamefont {Mallet}, \citenamefont {Gonz\'alez-Herrero},
  \citenamefont {Trambly~de Laissardi\`ere}, \citenamefont {Ugeda},
  \citenamefont {Magaud}, \citenamefont {G\'omez-Rodr\'{\i}guez}, \citenamefont
  {Yndur\'ain},\ and\ \citenamefont {Veuillen}}]{PhysRevLett.109.196802}%
  \BibitemOpen
  \bibfield  {author} {\bibinfo {author} {\bibfnamefont {I.}~\bibnamefont
  {Brihuega}}, \bibinfo {author} {\bibfnamefont {P.}~\bibnamefont {Mallet}},
  \bibinfo {author} {\bibfnamefont {H.}~\bibnamefont {Gonz\'alez-Herrero}},
  \bibinfo {author} {\bibfnamefont {G.}~\bibnamefont {Trambly~de
  Laissardi\`ere}}, \bibinfo {author} {\bibfnamefont {M.~M.}\ \bibnamefont
  {Ugeda}}, \bibinfo {author} {\bibfnamefont {L.}~\bibnamefont {Magaud}},
  \bibinfo {author} {\bibfnamefont {J.~M.}\ \bibnamefont
  {G\'omez-Rodr\'{\i}guez}}, \bibinfo {author} {\bibfnamefont {F.}~\bibnamefont
  {Yndur\'ain}}, \ and\ \bibinfo {author} {\bibfnamefont {J.-Y.}\ \bibnamefont
  {Veuillen}},\ }\href {\doibase 10.1103/PhysRevLett.109.196802} {\bibfield
  {journal} {\bibinfo  {journal} {Phys. Rev. Lett.}\ }\textbf {\bibinfo
  {volume} {109}},\ \bibinfo {pages} {196802} (\bibinfo {year}
  {2012})}\BibitemShut {NoStop}%
\bibitem [{\citenamefont {Wei\ss{}e}\ \emph {et~al.}(2006)\citenamefont
  {Wei\ss{}e}, \citenamefont {Wellein}, \citenamefont {Alvermann},\ and\
  \citenamefont {Fehske}}]{RevModPhys.78.275}%
  \BibitemOpen
  \bibfield  {author} {\bibinfo {author} {\bibfnamefont {A.}~\bibnamefont
  {Wei\ss{}e}}, \bibinfo {author} {\bibfnamefont {G.}~\bibnamefont {Wellein}},
  \bibinfo {author} {\bibfnamefont {A.}~\bibnamefont {Alvermann}}, \ and\
  \bibinfo {author} {\bibfnamefont {H.}~\bibnamefont {Fehske}},\ }\href
  {\doibase 10.1103/RevModPhys.78.275} {\bibfield  {journal} {\bibinfo
  {journal} {Rev. Mod. Phys.}\ }\textbf {\bibinfo {volume} {78}},\ \bibinfo
  {pages} {275} (\bibinfo {year} {2006})}\BibitemShut {NoStop}%
\bibitem [{\citenamefont {Akzyanov}\ \emph {et~al.}(2014)\citenamefont
  {Akzyanov}, \citenamefont {Sboychakov}, \citenamefont {Rozhkov},
  \citenamefont {Rakhmanov},\ and\ \citenamefont {Nori}}]{PhysRevB.90.155415}%
  \BibitemOpen
  \bibfield  {author} {\bibinfo {author} {\bibfnamefont {R.~S.}\ \bibnamefont
  {Akzyanov}}, \bibinfo {author} {\bibfnamefont {A.~O.}\ \bibnamefont
  {Sboychakov}}, \bibinfo {author} {\bibfnamefont {A.~V.}\ \bibnamefont
  {Rozhkov}}, \bibinfo {author} {\bibfnamefont {A.~L.}\ \bibnamefont
  {Rakhmanov}}, \ and\ \bibinfo {author} {\bibfnamefont {F.}~\bibnamefont
  {Nori}},\ }\href {\doibase 10.1103/PhysRevB.90.155415} {\bibfield  {journal}
  {\bibinfo  {journal} {Phys. Rev. B}\ }\textbf {\bibinfo {volume} {90}},\
  \bibinfo {pages} {155415} (\bibinfo {year} {2014})}\BibitemShut {NoStop}%
\bibitem [{\citenamefont {Colom\'es}\ and\ \citenamefont
  {Franz}(2018)}]{PhysRevLett.120.086603}%
  \BibitemOpen
  \bibfield  {author} {\bibinfo {author} {\bibfnamefont {E.}~\bibnamefont
  {Colom\'es}}\ and\ \bibinfo {author} {\bibfnamefont {M.}~\bibnamefont
  {Franz}},\ }\href {\doibase 10.1103/PhysRevLett.120.086603} {\bibfield
  {journal} {\bibinfo  {journal} {Phys. Rev. Lett.}\ }\textbf {\bibinfo
  {volume} {120}},\ \bibinfo {pages} {086603} (\bibinfo {year}
  {2018})}\BibitemShut {NoStop}%
\bibitem [{\citenamefont {Anderson}(1973)}]{anderson1973resonating}%
  \BibitemOpen
  \bibfield  {author} {\bibinfo {author} {\bibfnamefont {P.}~\bibnamefont
  {Anderson}},\ }\href {\doibase 10.1016/0025-5408(73)90167-0} {\bibfield
  {journal} {\bibinfo  {journal} {Materials Research Bulletin}\ }\textbf
  {\bibinfo {volume} {8}},\ \bibinfo {pages} {153} (\bibinfo {year}
  {1973})}\BibitemShut {NoStop}%
\bibitem [{\citenamefont {Sun}\ \emph {et~al.}(2010{\natexlab{b}})\citenamefont
  {Sun}, \citenamefont {Fertig},\ and\ \citenamefont {Brey}}]{sun2010}%
  \BibitemOpen
  \bibfield  {author} {\bibinfo {author} {\bibfnamefont {J.}~\bibnamefont
  {Sun}}, \bibinfo {author} {\bibfnamefont {H.~A.}\ \bibnamefont {Fertig}}, \
  and\ \bibinfo {author} {\bibfnamefont {L.}~\bibnamefont {Brey}},\ }\href
  {\doibase 10.1103/PhysRevLett.105.156801} {\bibfield  {journal} {\bibinfo
  {journal} {Phys. Rev. Lett.}\ }\textbf {\bibinfo {volume} {105}},\ \bibinfo
  {pages} {156801} (\bibinfo {year} {2010}{\natexlab{b}})}\BibitemShut
  {NoStop}%
\bibitem [{\citenamefont {Ando}(2005{\natexlab{b}})}]{ando2005}%
  \BibitemOpen
  \bibfield  {author} {\bibinfo {author} {\bibfnamefont {T.}~\bibnamefont
  {Ando}},\ }\href@noop {} {\bibfield  {journal} {\bibinfo  {journal} {Journal
  of the Physical Society of Japan}\ }\textbf {\bibinfo {volume} {74}},\
  \bibinfo {pages} {777} (\bibinfo {year} {2005}{\natexlab{b}})}\BibitemShut
  {NoStop}%
\end{thebibliography}%

\end{document}